\documentclass[a4paper,fleqn]{cas-sc}

\usepackage[numbers]{natbib}
\usepackage{subcaption}
\usepackage{rotating}
\usepackage{multirow}
\usepackage[ruled,vlined]{algorithm2e} 
\SetArgSty{textnormal}

\usepackage{amsmath} 
\usepackage{amssymb} 
\usepackage{amsfonts}
\usepackage{amsxtra}
\usepackage{tensor}
\usepackage{ulem}

\DeclareMathOperator*{\argmin}{\arg\!\min}
\newcommand{\tr}{\mathrm{tr}}
\newcommand{\diagm}{\mathrm{diag}}
\newcommand{\approximate}[1]{\tilde{#1}}

\newcommand{\fourier}[1]{\overline{#1}}
\newcommand{\laufvar}[1]{\breve{#1}}
\newcommand{\complex}[1]{\left(#1\right)^{*}}

\newcommand{\norm}[1]{\left|#1\right|}

\newcommand{\optsuperscript}{\text{opt}}
\newcommand{\elesuperscript}{\prime}
\newcommand{\ellipsoidsuperscript}{\mathrm{ellipsoid}}
\newcommand{\rvesuperscript}{\mathrm{RVE}}
\newcommand{\rveellipsuperscript}{\mathrm{RVE,e}}

\newcommand{\euler}{\mathrm{e}}

\newcommand{\ve}[1]{\boldsymbol{#1}} 
\newcommand{\te}[1]{\boldsymbol #1} 

\newcommand{\M}[1]{\uuline{\boldsymbol #1}} 
\newcommand{\V}[1]{\uline{\boldsymbol #1}} 

\newcommand{\semiax}{a}

\newcommand{\shapmat}{A}

\newcommand{\distchar}{f}
\newcommand{\cumdistchar}{\expandafter\uppercase\expandafter{\distchar}}

\newcommand{\eleangle}{\alpha}
\newcommand{\unitvec}[1]{\ve{e}_{#1}}
\newcommand{\fffE}{E}
\newcommand{\sffe}{e}
\newcommand{\fffF}{F}
\newcommand{\sfff}{f}
\newcommand{\auxmeancurve}[1]{f_{#1}}
\newcommand{\fffG}{G}
\newcommand{\sffg}{g}
\newcommand{\ellipE}{\mathcal{E}}
\newcommand{\ellipF}{\mathcal{F}}

\newcommand{\covbasisvec}[1]{\ve{g}_{#1}}

\newcommand{\meancurve}{H}
\newcommand{\ellipsoiddof}{\V{\semiax},\V{\eleangle},\ve{\elepos}}
\newcommand{\imagunit}{\mathrm{i}\mkern1mu}
\newcommand{\indicator}{\mathcal{I}}

\newcommand{\fourierindicator}[1]{\fourier{{\indicator}}_{#1}}

\newcommand{\lossfun}{\mathcal{L}}

\newcommand{\lengthrve}{l}
\newcommand{\body}{\Omega}

\newcommand{\normal}{n}

\newcommand{\nfourier}[1]{\fourier{N}_{#1}}

\newcommand{\samplesuperscript}{\mathrm{incl}}
\newcommand{\nincl}{N^{\samplesuperscript}}

\newcommand{\elepos}{p}

\newcommand{\relpos}{r}
\newcommand{\euclidr}{\mathbb{R}^3}
\newcommand{\rotmat}{R}
\newcommand{\surfele}[1]{\diff S}

\newcommand{\tpc}{S_2}
\newcommand{\generaltpc}[2]{\tpc^{#1 \rightarrow #2}}
\newcommand{\fouriertpc}[1]{\fourier{{\tpc}}_{#1}}
\newcommand{\elsymfun}[1]{\mathcal{S}_{#1}}

\newcommand{\vol}{V}

\newcommand{\vrve}{\vol^{\rvesuperscript}}
\newcommand{\minkowski}[3]{ \mathcal{W}^{#1,#2}_#3 }
\newcommand{\minkowskiellip}[3]{ {W}^{#1,#2}_#3 }

\newcommand{\minkowskiraw}[3]{{\minkowskiellip{#1}{#2}{#3}}^{\elesuperscript}}
\newcommand{\position}{x}
\newcommand{\spherex}{\hat{\position}}

\newcommand{\dofsrve}{\gamma}

\newcommand{\azimutangle}{\theta}
\newcommand{\polarangle}{\varphi}
\newcommand{\princurve}[1]{\kappa_{#1}}
\newcommand{\freqvec}{\omega}
\newcommand{\rvespace}{\Omega^{\rvesuperscript}}
\newcommand{\incspace}{\Omega^{\samplesuperscript}}
\newcommand{\incspaceellip}{\Omega^{\samplesuperscript\text{,e}}}
\newcommand{\unitsphere}{\Omega^{\mathrm{sphere}}}
\newcommand{\ellipsoid}{\Omega^{\ellipsoidsuperscript}}

\newcommand{\surfellipsoidprinc}{\partial{\Omega^{\ellipsoidsuperscript}}^{\elesuperscript}}

\newcommand{\diff}{\mathop{}\!\mathrm{d}}

    \def\tsc#1{\csdef{#1}{\textsc{\lowercase{#1}}\xspace}}
\tsc{WGM}
\tsc{QE}

\begin{document}
\normalem
\let\WriteBookmarks\relax
\def\floatpagepagefraction{1}
\def\textpagefraction{.001}

\shorttitle{Fast reconstruction of microstructures with ellipsoidal inclusions}    

\shortauthors{Seibert~et~al.}  

\title [mode = title]{Fast reconstruction of microstructures with ellipsoidal inclusions using analytical descriptors}  

\author[1]{Paul Seibert}[orcid=0000-0002-8774-8462] 
\credit{Conceptualization, Data Curation, Formal Analysis, Investigation, Methodology, Supervision, Validation, Visualization, Writing - original draft, Writing - review and editing} 
\cormark[1]

\author[1]{Markus Husert}[]
\credit{Investigation, Data Curation, Formal Analysis, Investigation, Methodology, Software, Validation, Visualization, Writing - original draft, Writing - review and editing}
\cormark[1]

\author[2]{Maximilian P. Wollner}[]
\credit{Conceptualization, Formal Analysis, Investigation, Methodology, Software, Supervision, Writing - review and editing}

\author[1]{Karl A. Kalina}[orcid=0000-0001-6170-4069]
\credit{Conceptualization, Methodology, Supervision, Writing - original draft, Writing - review and editing} 

\author[1,3]{Markus Kästner}[orcid=0000-0003-3358-1545]
\credit{Conceptualization, Funding acquisition, Resources, Writing - review and editing} 
\cormark[2]

\affiliation[1]{organization={Institute of Solid Mechanics},
    addressline={TU Dresden}, 
    city={Dresden},
    postcode={01069}, 
    country={Germany}}
\affiliation[2]{organization={Institute of Biomechanics},
    addressline={Graz University of Technology}, 
    city={Graz},
    postcode={8010}, 
    country={Austria}}
\affiliation[3]{organization={Dresden Center for Computational Materials Science},
    addressline={TU Dresden}, 
    city={Dresden},
    postcode={01069}, 
    country={Germany}}

\cortext[cor1]{Contributed equally}
\cortext[cor2]{Corresponding author}


\begin{abstract}
  Microstructure reconstruction is an important and emerging aspect of computational materials engineering and multiscale modeling and simulation.
  Despite extensive research and fast progress in the field, the application of descriptor-based reconstruction remains limited by computational resources.
  Common methods for increasing the computational feasibility of descriptor-based microstructure reconstruction lie in approximating the microstructure by simple geometrical shapes and by utilizing differentiable descriptors to enable gradient-based optimization.
  The present work combines these two ideas for structures composed of non-overlapping ellipsoidal inclusions such as magnetorheological elastomers.
  This requires to express the descriptors as a function of the microstructure parametrization.
  Deriving these relations leads to analytical solutions that further speed up the reconstruction procedure.
  Based on these descriptors, microstructure reconstruction is formulated as a multi-stage optimization procedure.
  The developed algorithm is validated by means of different numerical experiments and advantages and limitations are discussed in detail.
  \ \ 
\end{abstract}

\begin{graphicalabstract}
\includegraphics[width=0.7\textwidth]{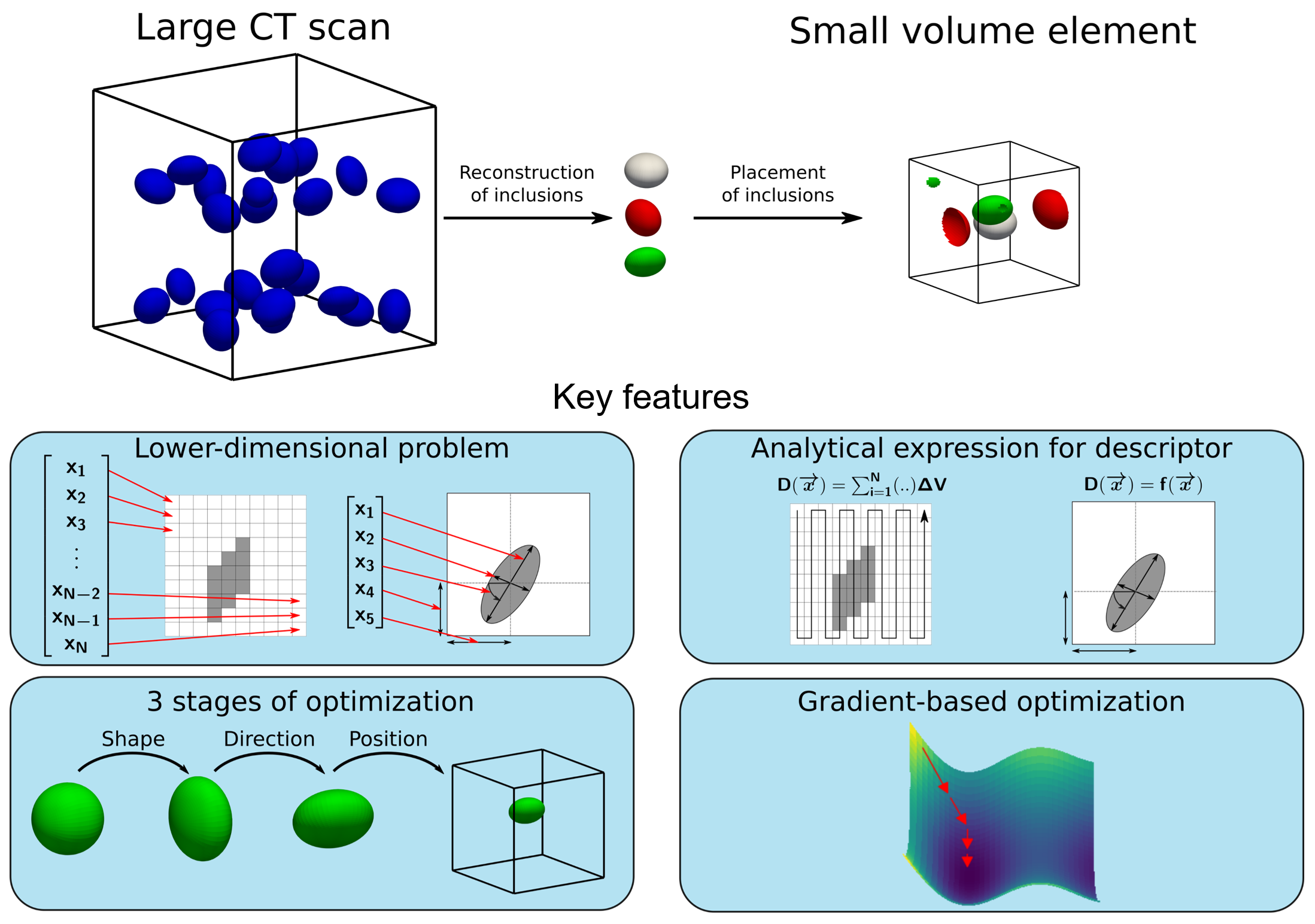}
\end{graphicalabstract}

\begin{highlights}
\item Analytical expressions for Minkoswki functionals and spatial correlations
\item Reconstruction from descriptors by efficient gradient-based optimization
\item Numerical experiments and discussion of limitations
\end{highlights}


\maketitle

\section{Introduction}
\label{sec:introduction}
Microstructure characterization and reconstruction (MCR) is an important and emerging field of research on the border between materials science, mechanics and computer science.
Harnessing recent hardware developments as well as new algorithms, it aims at accelerating materials engineering by enabling digital workflows such as numerical multiscale simulation~\cite{schroder_numerical_2014,vlassis_sobolev_2021,kalina_feann_2023} and computational (inverse) design~\cite{chen_data-centric_2022,choi_artificial_2022}.
This work presents a new MCR method tailored for materials that are composed of a matrix material with ellipsoidal inclusions.
Ellipsoids are often used to approximate inclusions, like in the seminal work by Eshelby~\cite{eshelby_elastic_1959}.
Furthermore, spherical or ellipsoidal particles are often embedded in an elastomer matrix in order to change the effective behavior~\cite{bergstrom_mechanical_1999}.
Examples comprise glass~\cite{eruslu_finite_2021}, silica~\cite{chen_microscopic_2021} or micron-sized magnetizable~\cite{hiptmair_design_2015} particles.
The latter are used in magnetorheological elastomers (MREs), which may serve as an exemplary motivation in this work. 
The microstructure of MREs can be specifically tuned during the production process.
For example, chain-like particle distributions~\cite{hiptmair_design_2015} or honeycombs~\cite{martin_generating_2003} can be generated.
In face of the high computational cost of magneto-mechanical and other simulations~\cite{mukherjee_microstructurally-guided_2019, kalina_multiscale_2020}, it is promising to use microstructure reconstruction for generating a small, periodic domain from a large, aperiodic volume like a computed tomography scan~\cite{schumann_microstructure_2021}.

A brief introduction on microstructure reconstruction techniques with a focus on recent techniques for random heterogeneous media is given in the following. 
The reader is referred to the reviews~\cite{bargmann_generation_2018,bostanabad_computational_2018,sahimi_reconstruction_2021} for a more detailed overview.
For this purpose, a distinction is made between \emph{descriptor-based} and \emph{data-based} methods. 
Despite exceptions~\cite{li_transfer_2018,robertson_local-global_2023}, most methods do not lie in a spectrum between these two extremes, but can clearly be classified into one of these categories.

In data-based methods, a generative model is fitted or trained on a sufficiently large data set of microstructures and is then used to sample new realizations of the same structure.
The model types are often inspired by texture synthesis~\cite{wei_state_2009} or generative models deep learning and range from local neighborhood-based methods~\cite{bostanabad_characterization_2016,latka_microstructure_2021,fu_hierarchical_2023} or generative adversarial networks~\cite{fokina_microstructure_2020,kench_generating_2021,henkes_three-dimensional_2022}, which are sometimes combined with autoencoders~\cite{zhang_slice--voxel_2021,zhang_da-vegan_2023}, to transformers~\cite{phan_size-invariant_2022} and the recent diffusion models~\cite{dureth_conditional_2023,lee_microstructure_2023}.
Much research is focused on identifying suitable model types and adapting them to microstructure reconstruction by enabling 2D-to-3D reconstruction~\cite{zhang_slice--voxel_2021,kench_generating_2021} making them applicable to small data sets~\cite{dureth_conditional_2023} or ensuring that certain descriptor requirements are met~\cite{li_transfer_2018,robertson_local-global_2023}.

Descriptor-based methods are not trained on a data set but are given a set of microstructure descriptors and corresponding values.
Volume fractions, spatial $n$-point correlations~\cite{yeong_reconstructing_1998}, the cluster correlation function~\cite{jiao_superior_2009}, entropic descriptors~\cite{piasecki_entropic_2010} and Minkowski functionals~\cite{schroder-turk_tensorial_2010} are typical candidates.
The method of choice depends on the application and two common examples are presented in \autoref{sec:methodsdescriptors}.
Based on these descriptors, microstructure reconstruction is often but not always~\cite{robertson_efficient_2021} formulated as an optimization problem in the space of possible microstructures.
For this purpose, the loss function of a microstructure is given by the deviation of the microstructure descriptors from their desired values.

The Yeong-Torquato algorithm~\cite{yeong_reconstructing_1998} is a very well-known descriptor-based reconstruction method, where the optimization problem is directly solved for pixel- or voxel-based microstructures using a specially adapted version of a common stochastic optimizer.
Unfortunately, the computational cost of this approach grows extremely at high resolutions and in 3D, where billions of iterations are needed~\cite{adam_efficient_2022}.
For this reason, several accelerated approaches have been presented~\cite{bostanabad_computational_2018}.
As an example, with a restricted choice of descriptors, efficient data structures can be employed to compute descriptor updates instead of recomputing the descriptor in every iteration~\cite{rozman_efficient_2001,adam_efficient_2022}.
Furthermore, the sampling rule and acceptance criterion of the pixel swap algorithms have been studied extensively~\cite{bostanabad_computational_2018} as well as reweighing the cost function~\cite{gerke_improving_2015} and introducing a multigrid scheme~\cite{pant_multigrid_2015,karsanina_hierarchical_2018,seibert_two-stage_2023}.
Two ideas that are considered in more detail in the following are (i) low-dimensional microstructure approximations and (ii) differentiable descriptors.

A common method of simplifying the optimization problem is to approximate the structure by simple geometric shapes such as spheres~\cite{horny_analysis_2022}.
Ellipsoids, for example, have been used by Xu et al.~\cite{xu_descriptor-based_2014} and Scheunemann et al~\cite{scheunemann_design_2015,scheunemann_scale-bridging_2017}.
For generic shapes, it is recently proposed to combine overlapping circles or spheres to represent complex 2D and 3D shapes~\cite{nakka_computationally_2022}.
In that work, however, the main focus lies on fitting all inclusions in a volume without overlapping, whereas descriptors are disregarded.
This is also a major difficulty with fiber composites, which are commonly approximated by rounded cylinders and are reconstructed from low-dimensional descriptors such as volume fraction and length distribution~\cite{schneider_sequential_2017,mehta_sequential_2022}.
Finally, for metallic materials, a structure is often created and parameterized by a Voronoi or Laguerre tessellation.
Examples include various codes such as \emph{DREAM.3D}~\cite{groeber_dream3d_2014}, \emph{Kanapy}~\cite{prasad_kanapy_2019}, \emph{NEPER}~\cite{quey_neperfepx_2022} and \emph{TOP}~\cite{kuhn_generating_2022}.
Generally, such microstructure approximations are beneficial because they immensely reduce the dimensionality of the optimization problem as compared to a pixel- or voxel-based microstructure representation.

Alternatively, without any approximations to the microstructure morphology, {differentiable} descriptors can be used to solve the underlying optimization problem using a gradient-based optimizer.
This has been presented as differentiable microstructure characterization and reconstruction (DMCR)~\cite{seibert_reconstructing_2021,seibert_descriptor-based_2022,seibert_two-stage_2023} and various approaches can be identified as special cases~\cite{li_transfer_2018,bostanabad_reconstruction_2020,bhaduri_efficient_2021}.
Furthermore, the publicly available \emph{MCRpy} package~\cite{seibert_microstructure_2022} implements DMCR as well as the Yeong-Torquato algorithm.

The central idea of the present work is that low-dimensional microstructure approximations and differentiable descriptors are not mutually exclusive but can be combined in an algorithm that is accelerated by both methods.
The difficulty lies in the fact that the descriptors need to be formulated in terms of and differentiable with respect to the low-dimensional microstructure parametrization. 
Deriving these relations for microstructures composed of non-overlapping ellipsoids is one central contribution of this work.
In fact, these derivations lead to semi-analytical solutions for the descriptors that are very efficient to compute, thus providing a third source of speedup.
These descriptors are used to formulate microstructure reconstruction as an optimization problem.
Similar to previous reconstruction algorithms~\cite{piasecki_two-stage_2020,robertson_local-global_2023,fu_hierarchical_2023,seibert_two-stage_2023} this optimization is divided into different stages for a further increase in efficiency.

The work is structured as follows.
The developed method is presented in \autoref{sec:methods}.
More precisely, the descriptors of choice are presented in \autoref{sec:methodsdescriptors} and the parametrization of the microstructures is laid out in \autoref{sec:methodsrepresentation}.
Analytical expressions of the descriptors are derived in \autoref{sec:methodsderivations}. 
This forms the basis for formulating microstructure reconstruction as a multi-stage gradient-based optimization procedure in \autoref{sec:methodsreconstruction}.
Numerical experiments are carried out and discussed in \autoref{sec:numericalexperiments} and the results are summarized in \autoref{sec:summary}.

The following notation is used.
Lower- and uppercase bold letters denote tensors of first ($\ve{x}$) and second ($\te{{\minkowski{0}{2}{2}}}$) order, respectively. 
In contrast, vectors ($\V{\eleangle}$) and matrices ($\M{\eleangle}$) are bold and single and double underlined, respectively.
Non-bold quantities are scalars or other symbols.
The Einstein summation convention is not used.

\section{Methods}
\label{sec:methods}
\begin{figure}[t]
    \centering
    \includegraphics[width=0.7\textwidth]{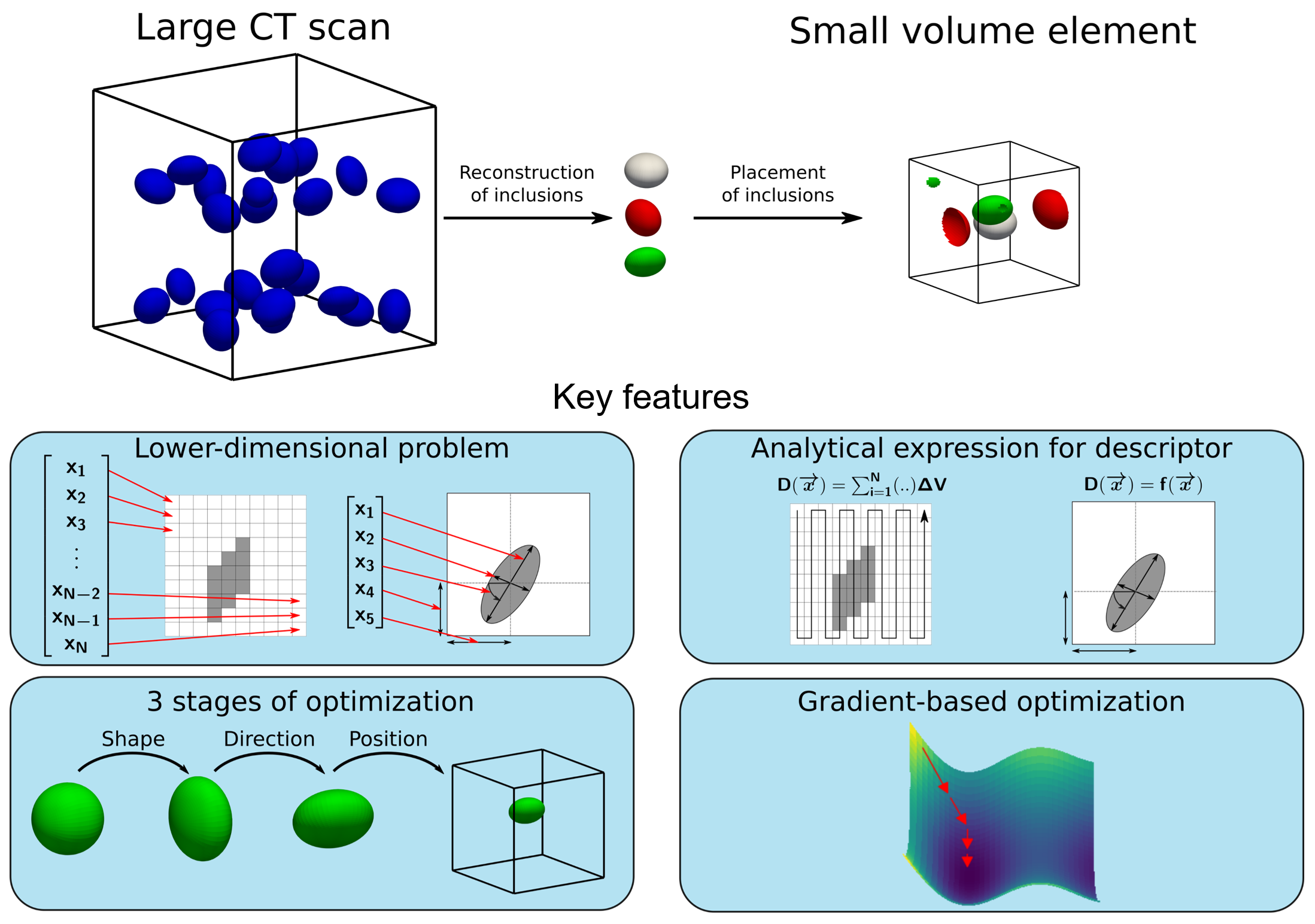}
    \caption{Schematic overview of the presented method. A (potentially large) 3D microstructure is characterized by the inclusions' Minkowski functionals and the spatial autocorrelation. From these descriptors, one or many statistical volume elements of arbitrary size can be reconstructed. The algorithm is rendered efficient by (i) the low dimensionality of the optimization problem, (ii) the analytical expressions for the descriptors, (iii) the division of the reconstruction process into multiple stages and (iv) the differentiability of the descriptors, enabling gradient-based optimization. \label{fig:schema}}
\end{figure}

An algorithm for the reconstruction of microstructures from given descriptors  is presented in the following.
These descriptors, namely the spatial two-point autocorrelation and the Minkowski functionals, are reviewed in \autoref{sec:methodsdescriptors}.
Very high efficiency is achieved by replacing the often used voxel-based microstructure representation, see \emph{MCRpy}~\cite{seibert_microstructure_2022}, by the degrees of freedom of ellipsoidal inclusions, see Scheunemann et al.~\cite{scheunemann_design_2015,scheunemann_scale-bridging_2017}.
This is outlined in \autoref{sec:methodsrepresentation}.
Unlike in previous works, these ellipsoids do not overlap.
Although distance computation and collision detection cause a significant computational overhead, this allows to directly reconstruct materials where the inclusions cannot overlap for physical reasons, such as magnetorheological elastomers.
Furthermore, this allows developing analytical solutions for microstructure descriptors as well as their derivatives, which greatly enhance the efficiency of the method.
This derivation is carried out in \autoref{sec:methodsderivations}.
Finally, based on the descriptor formulations, the reconstruction algorithm is formulated as an optimization problem in \autoref{sec:methodsreconstruction}.
This is summarized in \autoref{fig:schema}.

\subsection{Descriptors}
\label{sec:methodsdescriptors}
Minkowski functionals describe the shape of a smooth body in terms of integral values.
They are introduced in \autoref{sec:methodsdescriptorsminkowski} and are chosen in the present work for representing individual inclusions.
In contrast, the two-point correlation as introduced in \autoref{sec:methodsdescriptorstpc} captures information about the spatial distribution of a phase.
Here, this accounts for both the shape as well as the relative arrangement of the inclusions.
Since the former is already defined by the Minkowski functionals, we use the two-point correlation solely for the positioning of the ellipsoids.

\subsubsection{Minkowski functionals}
\label{sec:methodsdescriptorsminkowski}
Minkowski functionals are integral measures of the curvature of a smooth body, as such they describe the shape of the inclusions.
They can be tensors of arbitrary rank, even zero in the case of scalar functionals. According to Beisbart et al.~\cite{beisbart_vector-_2002} and Schröder-Turk et al.~\cite{schroder-turk_tensorial_2010} the general expressions of a Minkowski functional of an inclusion $\body$ are
\begin{align}
\minkowski{p}{q}{\nu}(\body)&=\frac{1}{\nu \binom{d}{\nu}} \int_{\partial \body}  \elsymfun{\nu-1}(\princurve{1},..,\princurve{d-1})
\underbrace{\ve{\position} \otimes...\otimes\ve{\position}}_{p\text{ times}}
\otimes 
\underbrace{\ve{\normal} \otimes...\otimes\ve{\normal}}_{q\text{ times}}
\surfele{d-1}\quad\text{and}\\
\minkowski{p}{0}{0}(\body)&=\int_{ \body} 
\underbrace{\ve{\position} \otimes...\otimes\ve{\position}}_{p\text{ times}}
\diff \ve{\position}
.
\end{align}
It should be noted that the Minkowski functional $\minkowski{p}{q}{\nu}$ is not formatted like a tensor, since its tensor rank depends on the values of $p$ and $q$ and it may also be a scalar quantity.
In the above equation, $p$, $q$ and $\nu$ are indices of the tensor, $d$ is the dimensionality of the inclusion $\body$, $\ve{\position}$ and $\ve{\normal}$ are the position and unit normal vector respectively, $\princurve{i}$ is the $i$th principal curvature of the surface, $\surfele{d-1}$ is an infinitesimal surface element and $\elsymfun{\nu}(\princurve{1},..,\princurve{d-1})$ is the elementary symmetric function, defined by Schneider~\cite{schneider_tensor_1999} for $\nu\in\{0,1,2\}$ as
\begin{align}
\label{eq:elsymfun}
{
\elsymfun{\nu}(\princurve{1},..,\princurve{d-1})=
\left\{\begin{aligned}\quad
&1,\quad &\text{if } \nu=0\\
\sum_{1\leq j \leq d-1} &\princurve{j},\quad &\text{if } \nu=1\\
\sum_{1\leq i \le j \leq d-1} &\princurve{i} \princurve{j},\quad &\text{if } \nu=2\\
\end{aligned}
\right.
} \; .
\end{align}

Herein, the scalar Minkowski functionals of interest are
\begin{align}
\label{eq:scal-mink1}
\minkowski{0}{0}{0}(\body) &=\int_{ \body} \diff \ve{\position},\\
\minkowski{0}{0}{1}(\body) &=\frac{1}{3} \int_{\partial \body}  \surfele{2} \quad\text{and}\\
\label{eq:scal-mink2}
\minkowski{0}{0}{2}(\body) &=\frac{1}{6} \int_{\partial \body} (\princurve{1} +\princurve{2}) \surfele{d-1},
\end{align}
where $\minkowski{0}{0}{0}$ coincides with the volume and $\minkowski{0}{0}{1}$ with one third of the surface area of the inclusion.\\
Among the tensorial functionals,
\begin{align}
\label{eq:tens-mink1}
\te{{\minkowski{0}{2}{1}}}(\body)&=\frac{1}{3} \int_{\partial \body}   \ve{\normal} \otimes \ve{\normal} \surfele{2} \quad\text{and} \\
\label{eq:tens-mink2}
\te{{\minkowski{0}{2}{2}}}(\body) &=\frac{1}{6} \int_{\partial \body}   \ve{\normal} \otimes \ve{\normal} (\princurve{1} +\princurve{2}) \surfele{2}
\end{align}
are relevant to this work.
They can be interpreted as the normal and curvature distribution, respectively, and more information is given in~\cite{schroder-turk_minkowski_2011}.
It should also be noted that the scalar Minkowski functionals $\minkowski{0}{0}{1}$ and $\minkowski{0}{0}{2}$ correspond to the traces of the tensorial functionals $\tr\left(\te{{\minkowski{0}{2}{1}}}\right)$ and $\tr\left(\te{{\minkowski{0}{2}{2}}}\right)$, respectively.

\subsubsection{Two-point correlation}
\label{sec:methodsdescriptorstpc}
The two-point correlation $\generaltpc{a}{b}(\ve{\relpos})$ describes the probability of finding the phases $a$ and $b$ at the start and end of the vector $\ve{\relpos}$, if it is placed randomly in the structure.
It can be shown that knowledge of one of the four possible two-point correlation, e.g. $\generaltpc{1}{1}$, uniquely determines the remaining three.~\cite{niezgoda_delineation_2008}.
For that reason, only the autocorrelation of the inclusion phase
\begin{equation}
\label{eq:defS2}
{
\tpc^{\rvesuperscript}(\ve{\relpos})= \generaltpc{1}{1} =
\frac{1}{\vrve}
\int_{\rvespace} \indicator(\ve{\position}) \cdot \indicator(\ve{\position}+\ve{\relpos}) \diff \ve{\position}
}
\end{equation}
is considered and hereafter referred to as the two-point correlation.
Herein, $\rvespace$ and $\vrve$ denote the domain and the volume of the RVE respectively and $\indicator^{\rvesuperscript}(\ve{\position})$ is the indicator-function
\begin{equation}
\indicator^{\rvesuperscript}(\ve{\position})=
\begin{cases}
1,\quad \text{if } \ve{\position} \in \incspace\\
0,\quad \text{otherwise} \\
\end{cases},
\end{equation}
where the inclusion domain  $\incspace$ is periodically extended beyond the RVE.
For this reason, $\indicator^{\rvesuperscript}(\ve{\position})$ and $\tpc^{\rvesuperscript}(\ve{\relpos})$ are also periodic and can be represented as Fourier series
\begin{equation}
\label{eq:fourierseries-S2}
{
\tpc^{\rvesuperscript}(\ve{\relpos})=
\sum_{j=-\infty}^{\infty}
\sum_{k=-\infty}^{\infty}
\sum_{l=-\infty}^{\infty}
\fouriertpc{jkl}^{\rvesuperscript} \euler^{\imagunit \ve{\freqvec}(j,k,l) \cdot \ve{\relpos}}
} \quad\text{and}
\end{equation}
\begin{equation}
\label{eq:fourierseries-indicator}
{
\indicator^{\rvesuperscript}(\ve{\position})=
\sum_{j=-\infty}^{\infty}
\sum_{k=-\infty}^{\infty}
\sum_{l=-\infty}^{\infty}
\fourierindicator{jkl}^{\rvesuperscript} \euler^{\imagunit \ve{\freqvec}(j,k,l) \cdot \ve{\position}}
},
\end{equation}
with the Fourier coefficients
\begin{equation}
\label{eq:fourier-S2}
{
\fouriertpc{jkl}^{\rvesuperscript} =\fouriertpc{}^{\rvesuperscript}(\ve{\freqvec}(j,k,l))=\frac{1}{\vrve} \int_{\rvespace} \tpc(\ve{\relpos}) \euler^{-\imagunit \ve{\freqvec} \cdot \ve{\relpos}} \diff \ve{\relpos}
} \quad\text{and}
\end{equation}
\begin{equation}
\label{eq:fourier-indicator}
{
\fourierindicator{jkl}^{\rvesuperscript} =\fourierindicator{}^{\rvesuperscript}(\ve{\freqvec}(j,k,l))=\frac{1}{\vrve} \int_{\rvespace} \indicator(\ve{\position}) \euler^{-\imagunit \ve{\freqvec} \cdot \ve{\position}} \diff \ve{\position}
}.
\end{equation}
In these equations, the frequency vector $\ve{\freqvec}$ is defined as
\begin{equation}
\label{eq:freqvec}
{
\iffalse
[\freqvec_i](i,j,k)=[j\frac{\pi}{\lengthrve_1},k\frac{\pi}{\lengthrve_2},l\frac{\pi}{\lengthrve_3}]^T
\else
\ve{\freqvec}(j,k,l)=2j\frac{\pi}{\lengthrve_1} \unitvec{1}+2k\frac{\pi}{\lengthrve_2} \unitvec{2}+2l\frac{\pi}{\lengthrve_3} \unitvec{3}
\fi
},
\end{equation}
where $\lengthrve_1$, $\lengthrve_2$ and $\lengthrve_3$ are the side lengths of the RVE.

In practice, the two-point correlation is approximated by truncating the Fourier series expansion after $\nfourier{1}$, $\nfourier{2}$ and $\nfourier{3}$ terms in the three spatial coordinates, respectively, i.e.
\begin{equation}
\label{eq:finite-fourier}
\approximate{\tpc}^{\rvesuperscript}(\ve{\relpos})
=
\sum_{j=-\nfourier{1}}^{\nfourier{1}}
\sum_{k=-\nfourier{2}}^{\nfourier{2}}
\sum_{l=-\nfourier{3}}^{\nfourier{3}}
\fouriertpc{jkl}^{\rvesuperscript} \euler^{\imagunit \ve{\freqvec}(j,k,l) \cdot \ve{\relpos}} \; .
\end{equation}

\subsection{Representation of ellipsoids and microstructures}
\label{sec:methodsrepresentation}
A single ellipsoid can be represented as
\begin{equation}
\ellipsoid(\ellipsoiddof)=
\{\ve{\position} \in \euclidr\, |\,
(\ve{\position}-\ve{\elepos}) \cdot
\underbrace{
\te{\rotmat}(\V{\eleangle}) \cdot
\te{\shapmat}^{\elesuperscript}(\V{\semiax})
 \cdot \te{\rotmat}(\V{\eleangle})^T
}_{\te{\shapmat}(\V{\semiax},\V{\eleangle})}
\cdot (\ve{\position}-\ve{\elepos})-1 \leq 0
\} \; .
\end{equation}
Herein, the semi-axes $\V{\semiax}=[\semiax_1,\,\semiax_2,\,\semiax_3]$ defining its size are sorted by magnitude\footnote{The following derivations assume inequality. In the case of equality, a small value can be added to ensure this condition. Alternatively, the special case could be investigated in a future work.}
\begin{equation}
\semiax_1 > \semiax_2 > \semiax_3
\end{equation}
and arranged in a diagonal matrix
\begin{equation}
[\shapmat^{\elesuperscript}_{ij}](\V{\semiax})=\diagm(\semiax_1^{-2},\semiax_2^{-2},\semiax_3^{-2}) = \begin{bmatrix}
    \semiax_1^{-2} & 0 & 0 \\
	0 & \semiax_2^{-2} & 0 \\
	0 & 0 & \semiax_3^{-2}
  \end{bmatrix}.
\end{equation}
Furthermore,~$\te{\rotmat}$ denotes a second order rotation tensor defined by
\begin{equation}
\te{\rotmat}(\V{\eleangle})=\te{\rotmat}^x(\eleangle_1) \cdot \te{\rotmat}^y(\eleangle_2) \cdot \te{\rotmat}^z(\eleangle_3) \; ,
\end{equation}
where the superscript of $\te{\rotmat}$ denotes the axis and the extrinsic Tait-Bryan angles $\V{\eleangle}=[\eleangle_1,\,\eleangle_2,\,\eleangle_3]$ define the rotation angle.
Finally, the position vector $\ve{\elepos}=\elepos_1 \unitvec{1}+\elepos_2 \unitvec{2}+\elepos_3 \unitvec{3}$ defines the center location.

With this representation of a single ellipsoid and an RVE defined as a rectangular domain
\begin{equation}
\rvespace=\{
\ve{\position} \in \euclidr\, |0\, \leq \ve{\position} \cdot \unitvec{1}< \lengthrve_{x} \text{ and } 0 \leq \ve{\position} \cdot \unitvec{2}< \lengthrve_{y} \text{ and } 0 \leq \ve{\position} \cdot \unitvec{3}< \lengthrve_{z}
\},
\end{equation}
a microstructure is defined by periodic continuation of all ellipsoids
\begin{equation}
\label{eq:inclusion-space-def}
\incspaceellip(\M{\semiax},\M{\eleangle},\M{\elepos})=\bigcup_{i=1}^{\nincl} \bigcup_{j,k,l=-\infty}^{\infty} \ellipsoid(\V{\semiax}^i,\V{\eleangle}^i,\ve{\elepos}^i+j\cdot\lengthrve_{x}\unitvec{1}+k\cdot\lengthrve_{y}\unitvec{2}+l\cdot\lengthrve_{z}\unitvec{3}) \; .
\end{equation}
Herein, a double underscore notation is introduced for~$\M{\semiax}$, $\M{\eleangle}$ and $\M{\elepos}$ to denote matrices where the $i$-th row contains the respective quantity for ellipsoid number~$i$.
Starting with the descriptors introduced in \autoref{sec:methodsdescriptors}, this notation allows analytical derivations that are presented in the following in \autoref{sec:methodsderivations}.

\subsection{Analytical derivations of descriptors for structures with ellipsoidal inclusions}
\label{sec:methodsderivations}
Minkowski functionals of single ellipsoids are derived in \autoref{sec:methodsderivationsminkowski}.
The arrangement of multiple ellipsoidal inclusions is described by the two-point autocorrelation, which is derived in \autoref{sec:methodsderivationstpc}.

\subsubsection{Minkowski functionals of ellipsoids}
\label{sec:methodsderivationsminkowski}
Because the considered Minkowski functionals are translation invariant, the ellipsoid is assumed to be located at the origin without loss of generality.
Moreover, the coordinate system is aligned with the principal axes of the ellipsoids and a back-transformation to the rotated coordinate system is applied to the Minkowski functionals after integration as
\begin{equation}
\te{{\minkowski{0}{2}{\nu}}}(\V{\semiax},\V{\eleangle})=\te{\rotmat}(\V{\eleangle}) \cdot \te{{\minkowskiraw{0}{2}{\nu}}}(\V{\semiax}) \cdot \te{\rotmat}(\V{\eleangle})^T.
\end{equation}
Due to symmetry reasons, the coordinates of $\te{{\minkowskiraw{0}{2}{\nu}}}$ form a diagonal matrix with three independent variables
\begin{equation}
[\minkowskiraw{0}{2}{\nu}_{jk}](\V{\semiax})=
\begin{bmatrix}
\minkowskiraw{0}{2}{\nu}_{11}(\V{\semiax}) & 0 & 0 \\
0 & \minkowskiraw{0}{2}{\nu}_{22}(\V{\semiax}) & 0 \\
0 & 0 & \minkowskiraw{0}{2}{\nu}_{33}(\V{\semiax})\\
\end{bmatrix}.
\end{equation}
Thus, the tensorial Minkowski functionals $\te{{\minkowski{0}{2}{\nu}}}$ for $\nu\in\{1,2\}$ are computed in the principal space by explicitly carrying out the integrals given in \autoref{sec:methodsdescriptorsminkowski}.

After this, the scalar Minkowski functionals can be obtained.
$\minkowski{0}{0}{0}(\V{\semiax})$ equals the volume of an ellipsoid,
\begin{equation}
\label{eq:volume-ellip-def}
\minkowski{0}{0}{0}(\V{\semiax})=\frac{4 \pi}{3} \semiax_1 \cdot \semiax_2 \cdot \semiax_3,
\end{equation}
and is easily defined by the product of its semi-axes.
Since the scalar Minkowski functionals 
\begin{equation}
\label{eq:scalar-mink-ellip-def}
\minkowski{0}{0}{\nu}(\V{\semiax})
=\tr\left(\te{{\minkowskiraw{0}{2}{\nu}}}(\V{\semiax})\right)
=
\sum_{i=1}^{3}
\minkowskiraw{0}{2}{\nu}_{ii}(\V{\semiax}),
\end{equation}
for $\nu\in\{1,2\}$, are the traces of their tensorial counterparts, they are rotation invariant and can be calculated as a function of the semi-axes only.

The remainder of this section completes the outlined derivation by carrying out the required integrals.
For this purpose, the surface of an unrotated ellipsoid at the coordinate origin $\surfellipsoidprinc(\V{\semiax})$ is defined in a parametric form by the function
\iffalse
\newcommand{\surfacepara}{(\azimutangle,\polarangle;\V{\semiax})}
\else
\newcommand{\surfacepara}{}
\fi
\begin{equation}
\label{eq:ellip-surface}
\ve{\position}(\azimutangle,\polarangle;\V{\semiax})=\semiax_1 \sin(\azimutangle) \cos(\polarangle) \unitvec{1}+\semiax_2 \sin(\azimutangle) \sin(\polarangle)\unitvec{2}+\semiax_3 \cos(\azimutangle)\unitvec{3},
\end{equation}
with the polar angle $0 \leq \azimutangle \leq \pi$ and the azimuth angle $0 \leq \polarangle \leq 2\pi$.

In order to calculate the tensorial Minkowski functionals of this surface, the expressions of the surface element~$\surfele{3}$, the unit normal vector~$\ve{\normal}$ and the mean curvature~$\meancurve$ are derived from the first and second fundamental form of differential geometry.
This is carried out explicitly in \autoref{sec:appendixderivations}.
The integrals in ~\autoref{eq:tens-mink1} and~\autoref{eq:tens-mink2} can be carried out and yield expressions of the tensorial Minkowski functionals of an unrotated ellipsoid
\begin{align}
\minkowskiraw{0}{2}{1}_{11}(\V{\semiax})&=\frac{2\pi \semiax_3^2}{3}\beta
\cdot
\left(
-\frac{1}{\gamma - \beta}
+\frac
{1}
{\sqrt{\beta \gamma (1-\beta)}}
\left[
\ellipF(\psi,\eta)+\frac{\beta}{\gamma-\beta}\ellipE(\psi,\eta)
\right]
\right),
\label{eq:first-minkowksi}
\\
\minkowskiraw{0}{2}{1}_{22}(\V{\semiax})&=\frac{2\pi \semiax_3^2}{3}\gamma
\cdot
\left(
\frac{1}{\gamma - \beta}
+\frac
{1}
{\sqrt{\beta \gamma (1-\beta)}(1-\gamma)}
\left[
\ellipF(\psi,\eta)-\frac{\gamma}{1-\eta}\ellipE(\psi,\eta)
\right]
\right),\\
\label{eq:133-minkowski}
\minkowskiraw{0}{2}{1}_{33}(\V{\semiax})&=\frac{2\pi \semiax_3^2}{3}
\frac{1}{\sqrt{\beta \gamma (1-\beta)}(1-\gamma)}
\left[
-\gamma \ellipF(\psi,\eta)+\ellipE(\psi,\eta)
\right],\\
\minkowskiraw{0}{2}{2}_{11}(\V{\semiax})&=\frac{\pi}{3} \semiax_3
\left(
\frac{\zeta}{\sqrt{\delta \zeta}}+
\frac{\delta\zeta+\delta-\zeta}
{(\delta-\zeta) \sqrt{\delta -1}}
\left[
\ellipF(\psi,\tau)-\ellipE(\psi,\tau)
\right]
\right),\\
\minkowskiraw{0}{2}{2}_{22}(\V{\semiax})&=\frac{\pi}{3(\zeta -1)} \semiax_3
\left(
-\frac{\zeta}{\sqrt{\delta \zeta}}+
\frac{\delta-\zeta-\delta \zeta}
{\tau \sqrt{\delta -1}}
\left[
(1-\tau) \ellipF(\psi,\tau)-\ellipE(\psi,\tau)
\right]
\right)\quad\text{and}\\
\minkowskiraw{0}{2}{2}_{33}(\V{\semiax})&=\frac{\pi}{3(\zeta -1)} \semiax_3
\left(
\frac{\zeta^2}{\sqrt{\delta \zeta}}+
\frac{\delta\zeta-\delta-\zeta}
{\sqrt{\delta -1}}
\ellipE(\psi,\tau)
\right),
\label{eq:last-minkowksi}
\end{align}
where the substitutions
\begin{align}
\beta&=\left(\frac{\semiax_3}{\semiax_1}\right)^2, \quad
\gamma=\left(\frac{\semiax_3}{\semiax_2}\right)^2,\quad
\delta=\left(\frac{\semiax_1}{\semiax_3}\right)^2,\quad
\zeta=\left(\frac{\semiax_2}{\semiax_3}\right)^2, \nonumber \\
\eta&=\frac{(1-\gamma)}{(1-\beta)},\quad
\tau=\frac{(\delta-\zeta)}{(\delta-1)} \quad \text{and} \quad
\psi=\arccos \left( \frac{\semiax_3}{\semiax_1} \right)
\end{align}
have been made for readability purposes.
Furthermore, $\ellipF$ and $\ellipE$ denote the incomplete elliptic integrals of the first and second kind, defined by Sidhu~\cite{sidhu_elliptic_1995} as
\begin{equation}
\ellipF(\psi,\eta)=\int_{0}^{\psi} \frac{1}{\sqrt{1-\eta^2 \sin^2(\laufvar{\psi})}} \diff \laufvar{\psi} \quad\text{and}\quad \ellipE(\psi,\eta)=\int_{0}^{\psi} \sqrt{1-\eta^2 \sin^2(\laufvar{\psi})} \diff \laufvar{\psi}.
\end{equation}

\subsubsection{Spatial correlations of ellipsoid distributions}
\label{sec:methodsderivationstpc}
The spatial autocorrelation can be computed from the indicator function of an RVE, which in turn is composed of indicator functions of individual non-overlapping ellipsoids.
Consequently, analytical solutions for the indicator function of single inclusions are derived in Fourier space.
These are then aggregated over all ellipsoids and used to compute the spatial autocorrelation.

After defining the indicator function of an ellipsoid $\indicator^{\ellipsoidsuperscript}$ as
\begin{equation}
\indicator^{\ellipsoidsuperscript}(\ve{\position};\ellipsoiddof)=
\begin{cases}
1,\quad \text{if } \ve{\position} \in \ellipsoid(\ellipsoiddof) \; , \\
0,\quad \text{otherwise}\; , \\
\end{cases}
\end{equation}
its Fourier transform follows as
\begin{align}
\fourierindicator{}^{\ellipsoidsuperscript}(\ve{\freqvec};\ellipsoiddof)&=
\frac{1}{\vrve} \int_{\rvespace} \indicator^{\ellipsoidsuperscript}(\ve{\position};\ellipsoiddof)
\euler^{-\imagunit \ve{\freqvec} \cdot \ve{\position}} \diff \ve{\position}
\nonumber\\
&=
\frac{1}{\vrve} \int_{\ellipsoid(\ellipsoiddof)} \euler^{-\imagunit \ve{\freqvec} \cdot \ve{\position}} \diff \ve{\position} \; .
\label{eq:fourierellipinit}
\end{align}
In order to solve the integral over the ellipsoid, it is mapped onto a unit sphere
\begin{equation}
\unitsphere=\{\ve{\position} \in \euclidr \,|\,\norm{\ve{\position}}\leq 1\}.
\end{equation}
This mapping takes the form of
\begin{equation}
\ve{\spherex}=\te{T} \cdot (\ve{\position}-\ve{\elepos}) \quad \text{and} \quad \ve{\position}= \te{T}^{-1} \cdot \ve{\spherex} + \ve{\elepos},
\end{equation}
with the transformation tensor
\begin{equation}
\te{T}(\V{\semiax},\V{\eleangle})= \sqrt{\te{\shapmat}^{\elesuperscript}}(\V{\semiax})\cdot \te{\rotmat}^T(\V{\eleangle})
\quad \text{with} \quad
[\sqrt{\shapmat^{\elesuperscript}}_{ij}](\V{\semiax})=\diagm(\semiax_1^{-1},\semiax_2^{-1},\semiax_3^{-1})
\end{equation}
and $\ve{\spherex}$ being the position vector after mapping onto the unit sphere.
Applying this transformation to \autoref{eq:fourierellipinit} yields
\begin{align}
\fourierindicator{}^{\ellipsoidsuperscript}(\ve{\freqvec};\ellipsoiddof)
&=
\frac{1}{\det(\te{T}) \vrve} \int_{\unitsphere} \euler^{-\imagunit \ve{\freqvec} \cdot (\te{T}^{-1} \cdot \ve{\spherex} + \ve{\elepos})} \diff \ve{\spherex}\nonumber \\
&=
\frac{\euler^{-\imagunit \ve{\freqvec} \ve{\elepos}}}{\det(\te{T}) \vrve} \int_{\unitsphere} \euler^{-\imagunit \ve{\freqvec} \cdot \te{T}^{-1} \cdot \ve{\spherex}} \diff \ve{\spherex}\nonumber \\
&=
\frac{\euler^{-\imagunit \ve{\freqvec} \ve{\elepos}}}{\det(\te{T}) \vrve} \int_{r=0}^1 \int_{\polarangle=0}^{\pi} \int_{\phi=0}^{2\pi}\euler^{-\imagunit \ve{\freqvec} \cdot \te{T}^{-1} \cdot \ve{\spherex}} r^2 \sin(\azimutangle) \diff \polarangle \diff \azimutangle \diff r \nonumber \\
&=
\frac{2 \pi \euler^{-\imagunit \ve{\freqvec}  \cdot  \ve{\elepos}}}{\det(\te{T}) \vrve} \int_{r=0}^1 \int_{\azimutangle=0}^{\pi} \euler^{-\imagunit r|\ve{\freqvec} \cdot \te{T}^{-1}| \cos(\azimutangle)} r^2 \sin(\azimutangle) \diff \azimutangle \diff r
.
\end{align}
By realizing that
\begin{equation}
\frac{\diff}{\diff \azimutangle} \euler^{-\imagunit r|\ve{\freqvec} \cdot \te{T}^{-1}| \cos(\azimutangle)}= \imagunit r \sin(\azimutangle)  |\ve{\freqvec} \cdot \te{T}^{-1}| \euler^{-\imagunit r |\ve{\freqvec} \cdot \te{T}^{-1}| \cos(\azimutangle)} ,
\end{equation}
the integral over $\azimutangle$ can be solved:
\begin{align}
\fourierindicator{}^{\ellipsoidsuperscript}(\ve{\freqvec};\ellipsoiddof)
&=
\frac{2\pi\euler^{-\imagunit \ve{\freqvec}  \cdot  \ve{\elepos}}}{\det(\te{T}) \vrve} \int_{r=0}^1
\left.
\euler^{-\imagunit r |\ve{\freqvec} \cdot \te{T}^{-1}| \cos(\azimutangle)}
\right|_{\azimutangle=0}^{\pi}
\frac{r}{\imagunit |\ve{\freqvec} \cdot \te{T}^{-1}|} \diff r\nonumber \\
&=
\frac{4 \pi \euler^{-\imagunit \ve{\freqvec}  \cdot  \ve{\elepos}}}{\det(\te{T}) \vrve}
\frac{\sin(|\ve{\freqvec} \cdot \te{T}^{-1}|) - |\ve{\freqvec} \cdot \te{T}^{-1}| \cos(|\ve{\freqvec} \cdot \te{T}^{-1}|)}
{|\ve{\freqvec} \cdot \te{T}^{-1}|^3} \\
\end{align}
By substituting
\begin{align}
\det(\te{T})(\V{\semiax})&=\frac{1}{\semiax_1 \semiax_2 \semiax_3}= \frac{4\pi}{3 \minkowskiellip{0}{0}{0}(\V{\semiax})}\quad \text{and}\\
|\ve{\freqvec} \cdot \te{T}^{-1}|&=
\sqrt{\ve{\freqvec} \cdot \te{T}^{-1} \cdot \te{T}^{-T} \cdot \ve{\freqvec}} \nonumber\\
&=
\sqrt{\ve{\freqvec} \cdot
\te{\shapmat}^{-1} \cdot \ve{\freqvec}} \; ,
\end{align}
the following solution is obtained:
\begin{align}
\label{eq:fourierellip}
\fourierindicator{jkl}^{\ellipsoidsuperscript}(\ellipsoiddof)&=\fourierindicator{}^{\ellipsoidsuperscript}(\ve{\freqvec}(j,k,l);\ellipsoiddof)=\frac{1}{\vrve} \int_{\ellipsoid(\ellipsoiddof)} \euler^{-\imagunit \ve{\freqvec} \cdot \ve{\position}} \diff \ve{\position}\nonumber \\
&=3\cdot \frac{\minkowskiellip{0}{0}{0}(\V{\semiax})}{\vrve} \cdot \euler^{-\imagunit \ve{\freqvec} \cdot  \ve{\elepos}} \cdot
\frac{
\sin(\Gamma)-\Gamma \cos(\Gamma)
}{\Gamma^3}
,
\end{align}
where $\Gamma=\sqrt{\ve{\freqvec} \cdot \te{\shapmat}^{-1}(\V{\semiax},\V{\eleangle}) \cdot \ve{\freqvec}}$.

The Fourier coefficients of an RVE containing $\nincl$ non-overlapping ellipsoids, $\fourierindicator{jkl}^{\rveellipsuperscript}$, can be calculated by summing the coefficients of  each ellipsoid due to the linearity of the Fourier transform,
\begin{align}
\label{eq:fourier-mult-ellip}
\fourierindicator{jkl}^{\rveellipsuperscript}(\M{\semiax},\M{\eleangle},\M{\elepos})&=\fourierindicator{}^{\rveellipsuperscript}(\ve{\freqvec}(j,k,l),(\M{\semiax},\M{\eleangle},\M{\elepos})) \nonumber \\
&=\sum_{i=1}^{\nincl} \fourierindicator{}^{\ellipsoidsuperscript}(\ve{\freqvec}(j,k,l),\V{\semiax}^i,\V{\eleangle}^i,\ve{\elepos}^i).
\end{align}
To obtain an expression for the autocorrelation in Fourier space $\fouriertpc{}^{\rvesuperscript}(\ve{\freqvec})$, \autoref{eq:defS2} and \autoref{eq:fourier-S2} are combined
\begin{align}
\fouriertpc{}^{\rvesuperscript}(\ve{\freqvec})&=\frac{1}{\vrve} \int_{\rvespace} \tpc^{\rvesuperscript}(\ve{\relpos}) \euler^{-\imagunit \ve{\freqvec} \cdot \ve{\relpos}} \diff \ve{\relpos}\nonumber \\
&=\frac{1}{(\vrve)^2} \int_{\rvespace}
\left[
\int_{\rvespace} \indicator^{\rvesuperscript}(\ve{\position}) \cdot \indicator^{\rvesuperscript}(\ve{\position}+\ve{\relpos}) \diff \ve{\position}
\right]
\euler^{-\imagunit \ve{\freqvec} \cdot \ve{\relpos}} \diff \ve{\relpos}\nonumber \\
&=\frac{1}{(\vrve)^2} \int_{\rvespace} \indicator^{\rvesuperscript}(\ve{\position})
\left[
\int_{\rvespace}  \indicator^{\rvesuperscript}(\ve{\position}+\ve{\relpos}) \euler^{-\imagunit \ve{\freqvec} \cdot \ve{\relpos}} \diff \ve{\relpos}
\right]
\diff \ve{\position} \; .
\end{align}
Substituting $\ve{\tilde{\position}}=\ve{\position}+\ve{\relpos}$ and applying the Wiener-Chintschin theorem yields
\begin{align}
\fouriertpc{}^{\rvesuperscript}(\ve{\freqvec})&=\frac{1}{(\vrve)^2} \int_{\rvespace} \indicator^{\rvesuperscript}(\ve{\position}) \euler^{\imagunit \ve{\freqvec} \cdot \ve{\position}}
\underbrace{
\left[
\int_{\rvespace}  \indicator^{\rvesuperscript}(\ve{\tilde{\position}}) \euler^{-\imagunit \ve{\freqvec} \cdot \ve{\tilde{\position}}} \diff \ve{\tilde{\position}}
\right]
}_{\vrve {\fourierindicator{}^{\rvesuperscript}(\ve{\freqvec})}}
\diff \ve{\position}\nonumber \\
&=\frac{{\fourierindicator{}^{\rvesuperscript}}}{\vrve}
\underbrace{
\int_{\rvespace} \indicator^{\rvesuperscript}(\ve{\position}) \euler^{\imagunit \ve{\freqvec} \cdot \ve{\position}}
\diff \ve{\position}
}_{\vrve \cdot \complex{\fourierindicator{}^{\rvesuperscript}(\ve{\freqvec})}}
\nonumber \\
&=\fourierindicator{}^{\rvesuperscript}(\ve{\freqvec}) \complex{\fourierindicator{}^{\rvesuperscript}(\ve{\freqvec})}=|\fourierindicator{}^{\rvesuperscript}(\ve{\freqvec})|^2
\end{align}
In the above derivation, $\complex{\bullet}$ denotes the complex conjugate. Also, it should be noted that the shift of the integration domain, which would be required by the substitution $\ve{\tilde{\position}}=\ve{\position}+\ve{\relpos}$, can be neglected since the integrant is periodic with a period length of exactly the RVE side length. For that reason, the above derivation is restricted to the frequency vectors defined in \autoref{eq:freqvec}.
The approximated two-point correlation of an RVE can then be expressed as
\begin{equation}
\approximate{\tpc}^{\rveellipsuperscript}(\ve{\relpos},\M{\semiax},\M{\eleangle},\M{\elepos})=
\sum_{j=-\nfourier{1}}^{\nfourier{1}}
\sum_{k=-\nfourier{2}}^{\nfourier{2}}
\sum_{l=-\nfourier{3}}^{\nfourier{3}}
\left|\fourierindicator{jkl}^{\rveellipsuperscript}(\M{\semiax},\M{\eleangle},\M{\elepos})\right|^2 \euler^{\imagunit \ve{\freqvec}(j,k,l) \cdot \ve{\relpos}} \; .
\label{eq:fouriertransforms2}
\end{equation}

\subsection{Reconstruction}
\label{sec:methodsreconstruction}
The previously derived expressions are used in the following to reconstruct microstructures given the desired descriptors.
Following the notation of Scheunemann et al.~\cite{scheunemann_design_2015,scheunemann_scale-bridging_2017}, $\V{\dofsrve}$ is used to denote the parametrization of the microstructure.
In the simple and direct case~\cite{yeong_reconstructing_1998,seibert_microstructure_2022},~$\V{\dofsrve}$ comprises all pixel or voxel values of the microstructure.
In the present work, however,~$\V{\dofsrve}$ is given by combining $\M\semiax$, $\M\eleangle$ and $\M\elepos$ as introduced in \autoref{sec:methodsrepresentation} in a single vector.
The optimal parametrization often is given by the solution of the optimization problem
\begin{equation}
\label{eq:optproblem}
\V{\dofsrve}^{\optsuperscript}=\argmin_{\V{\dofsrve}} \lossfun(\V{\dofsrve}) ,
\end{equation}
where the loss function~$\lossfun$ quantifies the difference between the current and desired descriptor values.
In the present case, the optimization problem is decomposed into a two-step procedure
\begin{enumerate}
    \item Reconstruct a set of ellipsoidal inclusions, where each ellipsoid is characterized by its semi-axes~$\V\semiax$ and orientations~$\V\eleangle$. The distribution of Minkowski functionals over all inclusions is used for this step.
    \item Place the inclusions in the domain under consideration of the spatial two-point autocorrelation. In this step, the ellipsoid shape and orientation remain fixed and only the position vectors are adjusted.
\end{enumerate}
This is shown in \autoref{alg:alg} and the individual stages are discussed in \autoref{sec:methodsrecincl} and \autoref{sec:methodsrecplace}, respectively.
While this approach is not necessarily optimal as given in \autoref{eq:optproblem}, it is highly efficient and yields very good results\footnote{As a future work, it might be promising to use the present results as an initialization for a convex optimizer for solving the full optimization problem in~\autoref{eq:optproblem}. }.
\begin{algorithm}[htb]
\DontPrintSemicolon
\SetAlgoLined
\KwIn{Set of original Minkowski functionals $\mathbb{W}^\mathrm{orig}$; volume fractions $v_\mathrm{f}$; desired autocorrelation $\approximate{\tpc}^{\mathrm{des}}(\ve{\relpos})$}
 $N^\mathrm{incl} =$ determine from $\mathbb{W}^\mathrm{orig}$ and $v_\mathrm{f}$   \tcp*{number of inclusions}
 \tcp*[l]{step 1 - reconstructed individual inclusions using Minkowski functionals}
 \For(\tcp*[f]{loop over inclusions}){$i = 1$ \KwTo $N^\mathrm{incl}$}{
  $\left(
    \minkowski{0}{0}{0}{}, \minkowski{0}{0}{1}{}, \minkowski{0}{0}{2}{},
    \te{{\minkowski{0}{2}{1}}}{},
    \te{{\minkowski{0}{2}{2}}}{} 
    \right)
    \sim 
    \mathbb{W}^\mathrm{orig}$  \tcp*{sample Minkowski functionals}
  $\ve{\semiax}^{\optsuperscript} = \argmin_{\ve{\semiax}} \lossfun^\mathrm{\semiax}(\V{\semiax})$ \tcp*{find semiaxes, use Eq. \eqref{eq:lossfunca}}
  $\ve{\eleangle}^{\optsuperscript} = \argmin_{\ve{\eleangle}} \lossfun^\mathrm{\eleangle}(\V{\semiax}^{\optsuperscript}, \V{\eleangle})$ \tcp*{find orientation, use Eq. \eqref{eq:lossfuncalpha}}
 }
 $\M{\semiax}^{\optsuperscript}, \, \M{\eleangle}^{\optsuperscript} = \bigsqcup_{i=1}^{\nincl} \V{\semiax}^{\optsuperscript}_i, \, \V{\eleangle}^{\optsuperscript}_i$  \tcp*{assemble over all inclusions}
 \tcp*[l]{step 2 - place inclusions using autocorrelation}
 $\M{\elepos}^{\optsuperscript}=\argmin_{\M{\elepos}} \left[ \lossfun^\mathrm{p}(\M{\semiax}^{\optsuperscript},\M{\eleangle}^{\optsuperscript},\M{\elepos}) + 
\lambda \cdot 
\lossfun^\mathrm{c}(\M{\semiax}^{\optsuperscript},\M{\eleangle}^{\optsuperscript},\M{\elepos}) \right]$ \tcp*{find positions, use Eq. \eqref{eq:lossfuncp} and \eqref{eq:lossfuncc}}
\KwOut{Reconstructed microstructure, parameterized by $\M{\semiax}^{\optsuperscript},\M{\eleangle}^{\optsuperscript}$ and $\M{\elepos}^{\optsuperscript}$}
 \caption{Proposed microstructure reconstruction algorithm.}
 \label{alg:alg}
\end{algorithm}

\subsubsection{Inclusion reconstruction}
\label{sec:methodsrecincl}
Given a set of Minkowski functionals of all inclusions in the original structure
\begin{equation}
    \mathbb{W}^\mathrm{orig} = \left\{ \left(
    \minkowski{0}{0}{0}{}^\mathrm{orig}, \minkowski{0}{0}{1}{}^\mathrm{orig}, \minkowski{0}{0}{2}{}^\mathrm{orig},
    \te{{\minkowski{0}{2}{1}}}{}^\mathrm{orig},
    \te{{\minkowski{0}{2}{2}}}{}^\mathrm{orig}
    \right)_i \right\}_{i=1}^{N_\mathrm{orig}^\mathrm{incl}}
\end{equation}
ellipsoids for an RVE to reconstruct can be sampled by repeatedly drawing tuples of Minkowski functionals until a sufficient number of ellipsoids is reached.
The number of ellipsoids is determined by the volume fraction.
If the reconstructed structure is of smaller or equal size as the original one, tuples of Minkowski functionals can be drawn without replacement, which is denoted by
\begin{equation}
    \left(
    \minkowski{0}{0}{0}{}, \minkowski{0}{0}{1}{}, \minkowski{0}{0}{2}{},
    \te{{\minkowski{0}{2}{1}}}{},
    \te{{\minkowski{0}{2}{2}}}{} 
    \right)
    \sim 
    \mathbb{W}^\mathrm{orig} \; .
\end{equation}
This ensures that the set of reconstructed inclusions~$\mathbb{W}^\mathrm{rec} \subseteq \mathbb{W}^\mathrm{orig}$.
If the original structure is significantly larger, advanced techniques such as stratified sampling may be used~\cite{olivier_uqpy_2020}.

After this sampling is completed, the semiaxes~$\V{\semiax}$ and orientation~$\V{\eleangle}$ need to be recovered from the Minkowski functionals.
Because the three scalar Minkowski functionals are rotation-invariant, they are used to determine~$\V{\semiax}$, whereas~$\V{\eleangle}$ follow from their tensorial counterparts.
For this purpose, two distinct optimization problems,
\begin{equation}
    \V{\semiax}^{\optsuperscript}=\argmin_{\V{\semiax}} \lossfun^\mathrm{\semiax}(\V{\semiax})
    \quad \mathrm{and} \quad
    \V{\eleangle}^{\optsuperscript}=\argmin_{\V{\eleangle}} \lossfun^\mathrm{\eleangle}(\V{\semiax}^{\optsuperscript}, \V{\eleangle}) \; ,
\end{equation}
are solved sequentially for each ellipsoid and the results are assembled to the matrices~$\M{\semiax}$ and~$\M{\eleangle}$.
The loss functions are chosen as
\begin{equation}
    \lossfun^\mathrm{\semiax}(\V{\semiax}) = \dfrac{1}{2} \left[ 
    \left( 
    \dfrac{{\minkowski{0}{0}{0}}(\V{\semiax})}{{\minkowski{0}{0}{0}}^\mathrm{des}} - 1
    \right)^2
    +
    \left( 
    \dfrac{{\minkowski{0}{0}{1}}(\V{\semiax})}{{\minkowski{0}{0}{1}}^\mathrm{des}} - 1
    \right)^2
    +
    \left( 
    \dfrac{{\minkowski{0}{0}{2}}(\V{\semiax})}{{\minkowski{0}{0}{2}}^\mathrm{des}} - 1
    \right)^2
    \right]
    \label{eq:lossfunca}
\end{equation}
and
\begin{equation}
    \lossfun^\mathrm{\eleangle}(\V{\semiax}, \V{\eleangle}) = \dfrac{1}{2} \cdot \tr \left[ 
    \left( 
    \dfrac{\te{{\minkowski{0}{2}{1}}}(\V{\semiax}, \V{\eleangle})}{\minkowski{0}{0}{1}(\V{\semiax})}  - \dfrac{\te{{\minkowski{0}{2}{1}}}{}^\mathrm{des}}{\minkowski{0}{0}{1}{}^\mathrm{des}}
    \right)^2
    +
    \left( 
    \dfrac{\te{{\minkowski{0}{2}{2}}}(\V{\semiax}, \V{\eleangle})}{\minkowski{0}{0}{2}(\V{\semiax})}  - \dfrac{\te{{\minkowski{0}{2}{2}}}{}^\mathrm{des}}{\minkowski{0}{0}{2}{}^\mathrm{des}}
    \right)^2
    \right] \; ,
    \label{eq:lossfuncalpha}
\end{equation}
where~$(\bullet)^2$ denotes an element-wise square when applied to a tensorial quantity.
The optimization problems are solved using the gradient-based BFGS optimizer~\cite{zhu_algorithm_1997}.
This stable and fast approach enables a solution in the order of milliseconds, even on a conventional laptop.

As an alternative to the presented sampling procedure, one feasible approach lies in defining a microstructure descriptor based on kernel density estimates (KDEs) of the Minkowski functionals over all ellipsoids.
Such a descriptor might be used for optimization instead of sampling.
Naturally, considering only KDEs of individual components neglects their correlations.
A simple example shows that this behavior is not desirable:
Consider an RVE with spherical inclusions of varying size.
Because the first two scalar Minkowski functionals~$\minkowski{0}{0}{0}$ and~$\minkowski{0}{0}{1}$ can be identified as the inclusion volume and surface area, respectively, reconstructing uncorrelated KDEs might lead to inclusions with low volume and high surface area, which can only be realized by flat or elongated ellipsoids. 
This limitation is discussed further in \autoref{sec:limitationsandscaling}.
As a solution to this problem, adding the covariance matrix to the descriptor or considering multi-dimensional KDEs might solve this problem.
However, these investigations exceed the scope of this work.

\subsubsection{Inclusion placing}
\label{sec:methodsrecplace}
After a representative set of inclusions, i.e., ellipsoid shapes and orientations, has been reconstructed from the Minkowski tensors, a volume element is created by placing these inclusions according to~$\tpc$.
This is expressed as a final optimization problem
\begin{equation}
\label{eq:optproblemseparated}
\M{\elepos}^{\optsuperscript}=\argmin_{\M{\elepos}} \left[ \lossfun^\mathrm{p}(\M{\semiax}^{\optsuperscript},\M{\eleangle}^{\optsuperscript},\M{\elepos}) + 
\lambda \cdot 
\lossfun^\mathrm{c}(\M{\semiax}^{\optsuperscript},\M{\eleangle}^{\optsuperscript},\M{\elepos}) \right]
\end{equation}
with a descriptor loss
\begin{equation}
    \lossfun^\mathrm{p}(\M{\semiax},\M{\eleangle},\M{\elepos}) = 
    \sum_{\ve{\relpos}} \left[
    \left(
    \approximate{\tpc}^{\rveellipsuperscript}(\ve{\relpos},\M{\semiax},\M{\eleangle},\M{\elepos}) -
    \approximate{\tpc}^{\mathrm{des}}(\ve{\relpos})
    \right)^2 
    \right]
    \label{eq:lossfuncp}
\end{equation}
formulated as a sum of squared errors in the spatial correlations and a contact penalty
\begin{equation}
    \lossfun^\mathrm{c}(\M{\semiax}^{\optsuperscript},\M{\eleangle}^{\optsuperscript},\M{\elepos}) = \sum_{i=1}^{N^\mathrm{incl}} \sum_{j=1}^{i-1} \left( \dfrac{\left\langle \delta^\mathrm{thr} - \delta(i, j) \right \rangle}{\min \M{\semiax}^{\optsuperscript}} \right)^{3} \; .
    \label{eq:lossfuncc}
\end{equation}
Herein,~$\langle {\bullet} \rangle$ denotes the Macauley brackets, also known as ReLU function. 
Furthermore, $\delta(i, j)$ denotes the distance\footnote{The distance is defined as the distance between the ellipsoid surfaces, not the centers. It is negative if the ellipsoids overlap and positive otherwise. Details on the distance computation are given by Perram et al.~\cite{perram_statistical_1985}.} between both ellipsoids, which is penalized if it exceeds the threshold $\delta^\mathrm{thr}$. 
In this work, $\delta^\mathrm{thr} = 0.1 \cdot \min \M{\semiax}^{\optsuperscript}$ is chosen.

Solving \autoref{eq:optproblemseparated} is the most computationally demanding part of the reconstruction process.
Unfortunately, simple convex minimization is observed to yield suboptimal local minima.
For this reason, a global optimization algorithm called Multi-Level Single-Linkage~\cite{rinnooy_kan_stochastic_1987,rinnooy_kan_stochastic_1987-1,kucherenko_application_2005} from the \emph{NLOpt} library~\cite{johnson_nlopt_2023} is used, where the low-storage variant of the gradient-based BFGS algorithm~\cite{zhu_algorithm_1997} serves as a local optimizer.
To provide a good initialization, ellipsoids are first placed individually in order of descending inclusion volume.
Without loss of generality, the largest inclusion placed at the center.

\section{Implementation and Results}
\label{sec:numericalexperiments}
The presented method is implemented in Julia and all numerical experiments are performed on a conventional laptop and only on a single thread.
It should be noted that the global optimization algorithm used in the ellipsoid placing step, which is the computational bottleneck of the method, in principle allows for straightforward parallelism.
Hence, significant efficiency gains are still achievable without changing the presented formulation merely by means of coding\footnote{These measures, however, do not affect the scalability of the method, but only reduce the wallclock time by a constant factor.}. 

In the remainder of this section, the method is validated and tested by means of various numerical experiments.
First, the reconstructed volume is set equally large as the original structure in order to validate the reconstruction and placement of spheres and ellipsoids in \autoref{sec:numericalexperiment1}.
In these cases, the algorithm is expected to exactly recover the original structure.
In contrast, this is not possible if the original structure is larger than the reconstructed domain.
Such cases are investigated in \autoref{sec:numericalexperiment3}.
Finally, the scalability and limitations of the method are discussed in \autoref{sec:limitationsandscaling}.

\subsection{Validation by same-size microstructures}
\label{sec:numericalexperiment1}
\begin{figure}[t]
	\centering
	\subfloat[Original three spheres structure]{\includegraphics[width=0.32\textwidth]{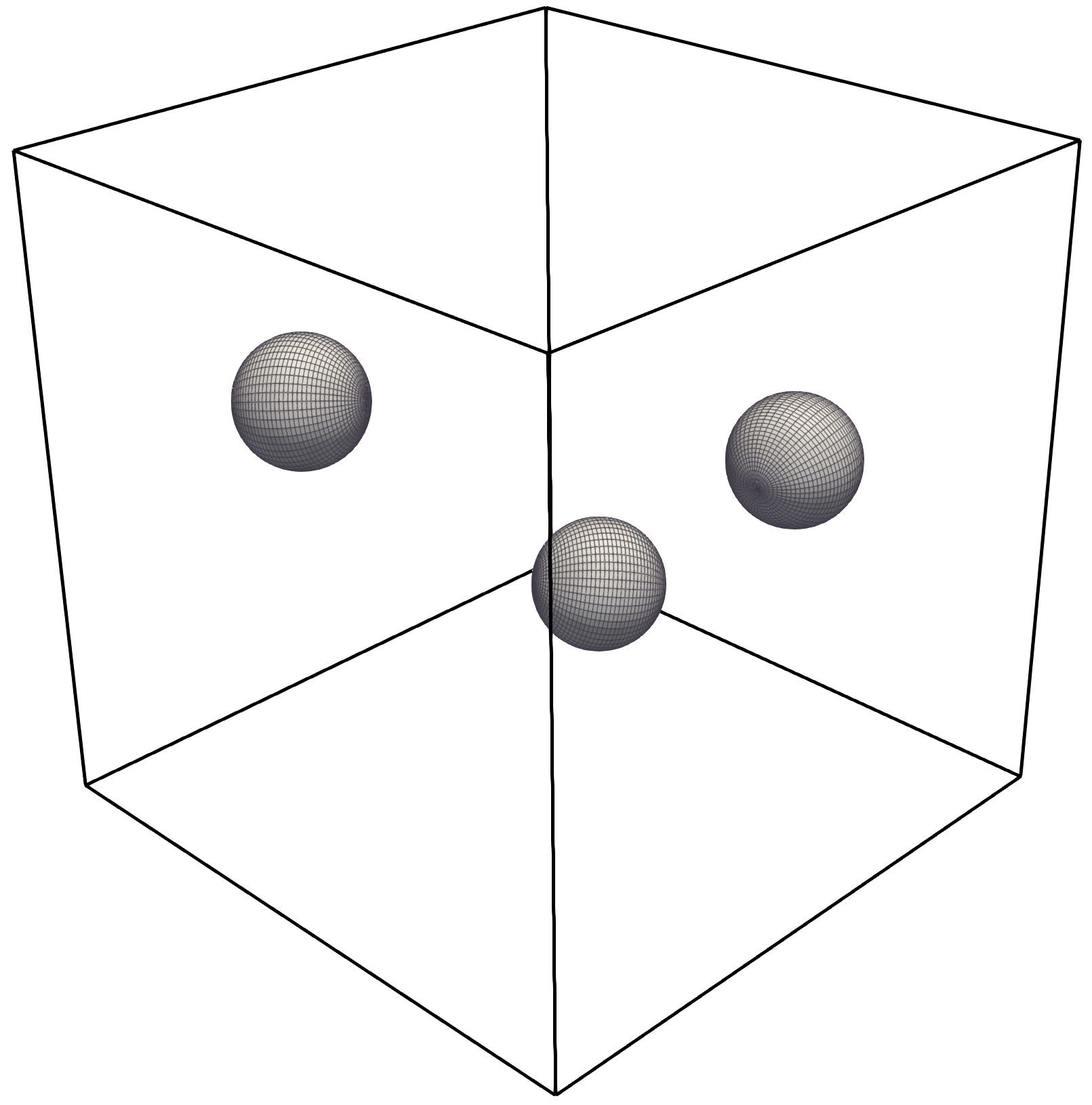}}
	\hfill
	\subfloat[Original three ellipsoids structure]{\includegraphics[width=0.32\textwidth]{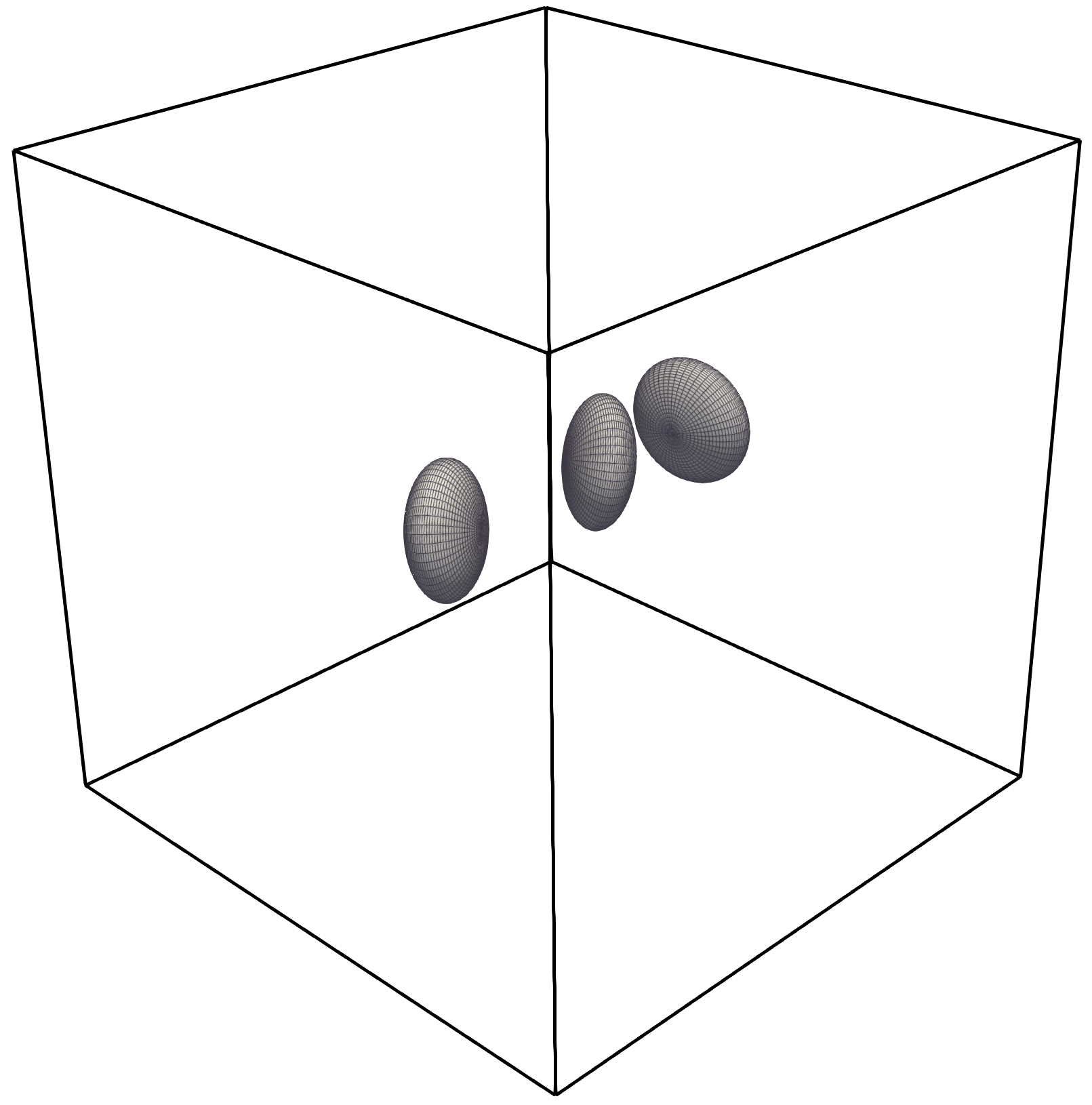}} 
	\hfill
	\subfloat[Original random spheres structure]{\includegraphics[width=0.32\textwidth]{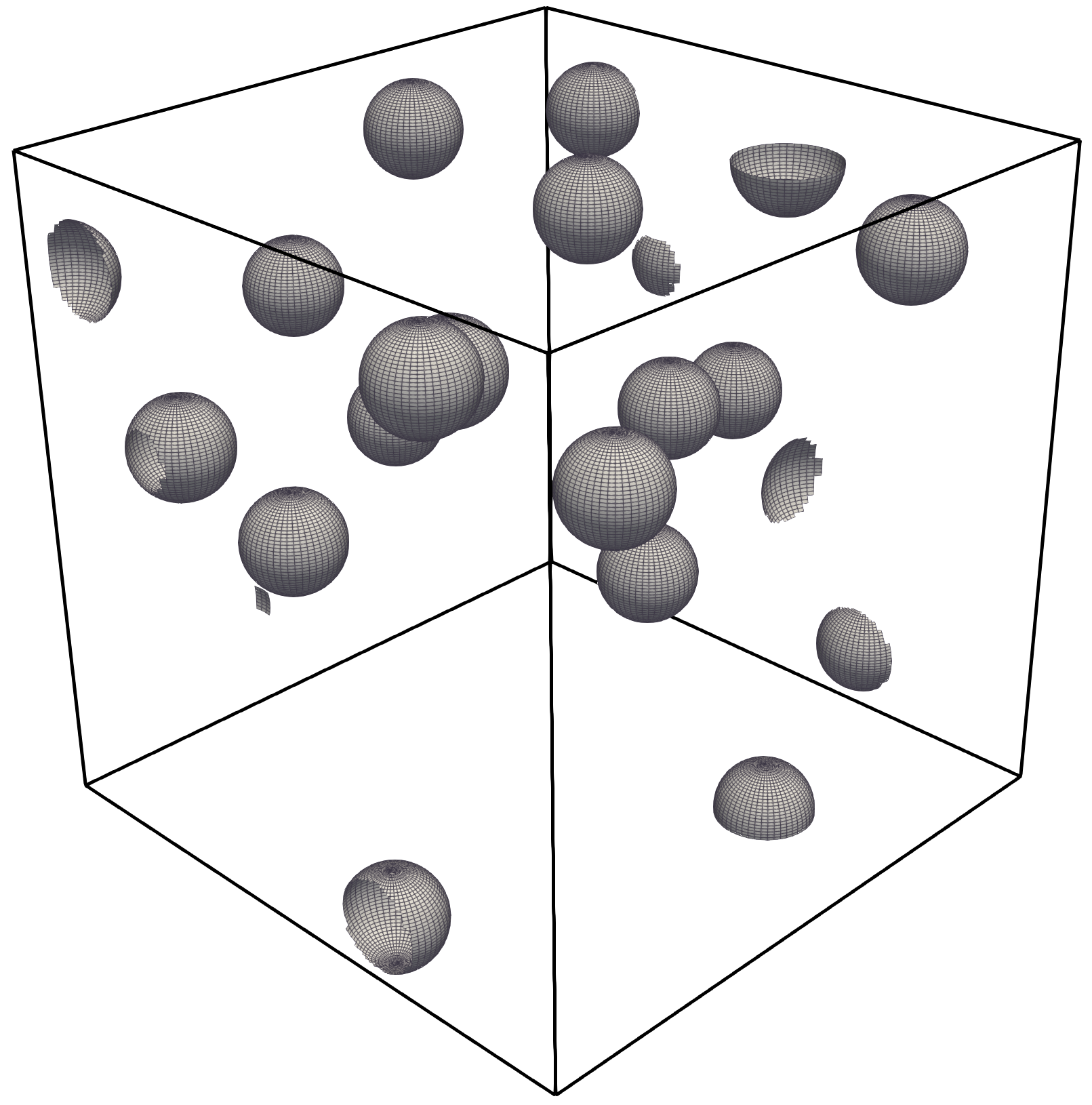}} 
	\vfill
	\subfloat[Reconstructed three spheres structure]{\includegraphics[width=0.32\textwidth]{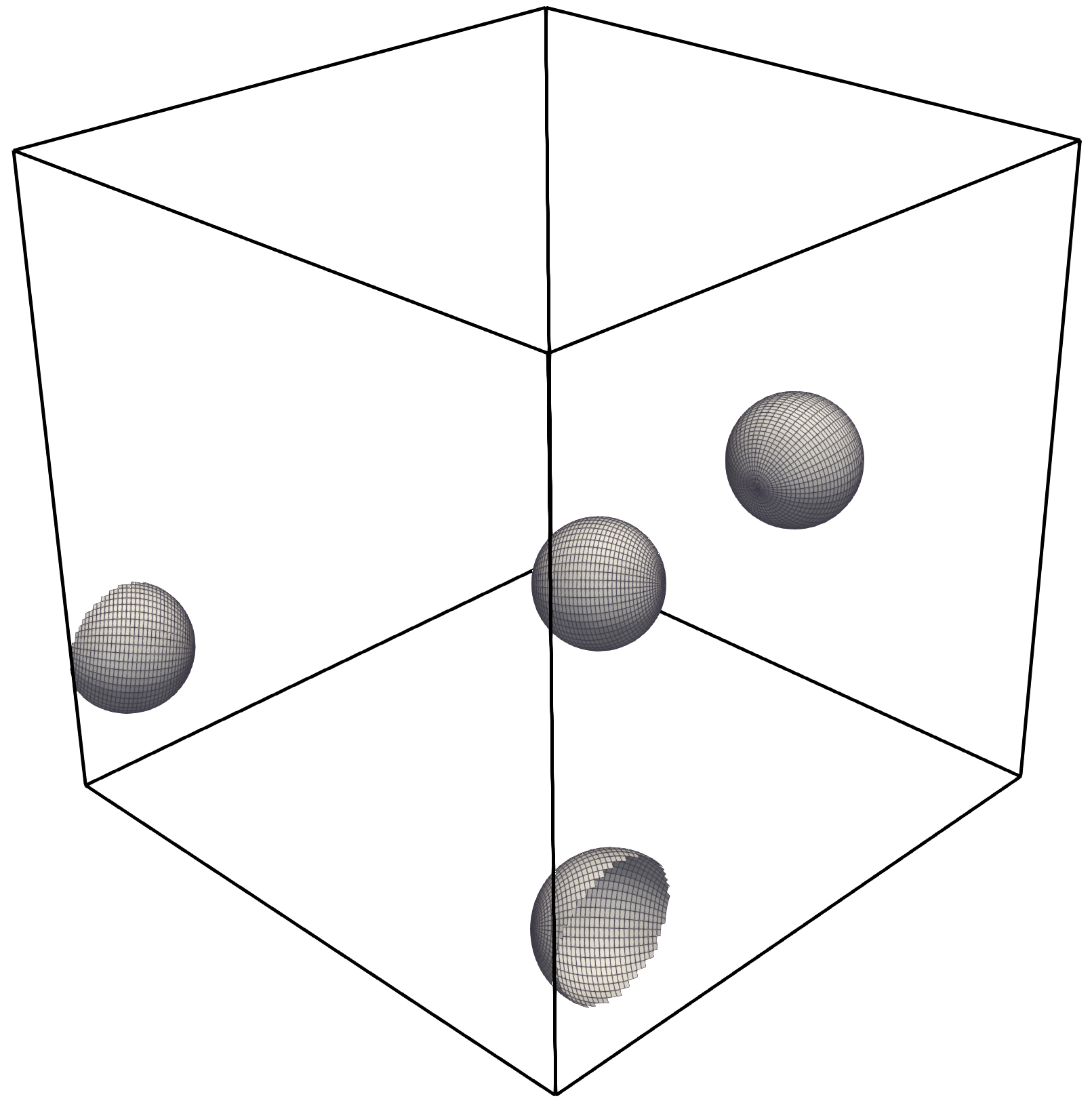}} 
	\hfill
	\subfloat[Reconstructed three ellipsoids structure]{\includegraphics[width=0.32\textwidth]{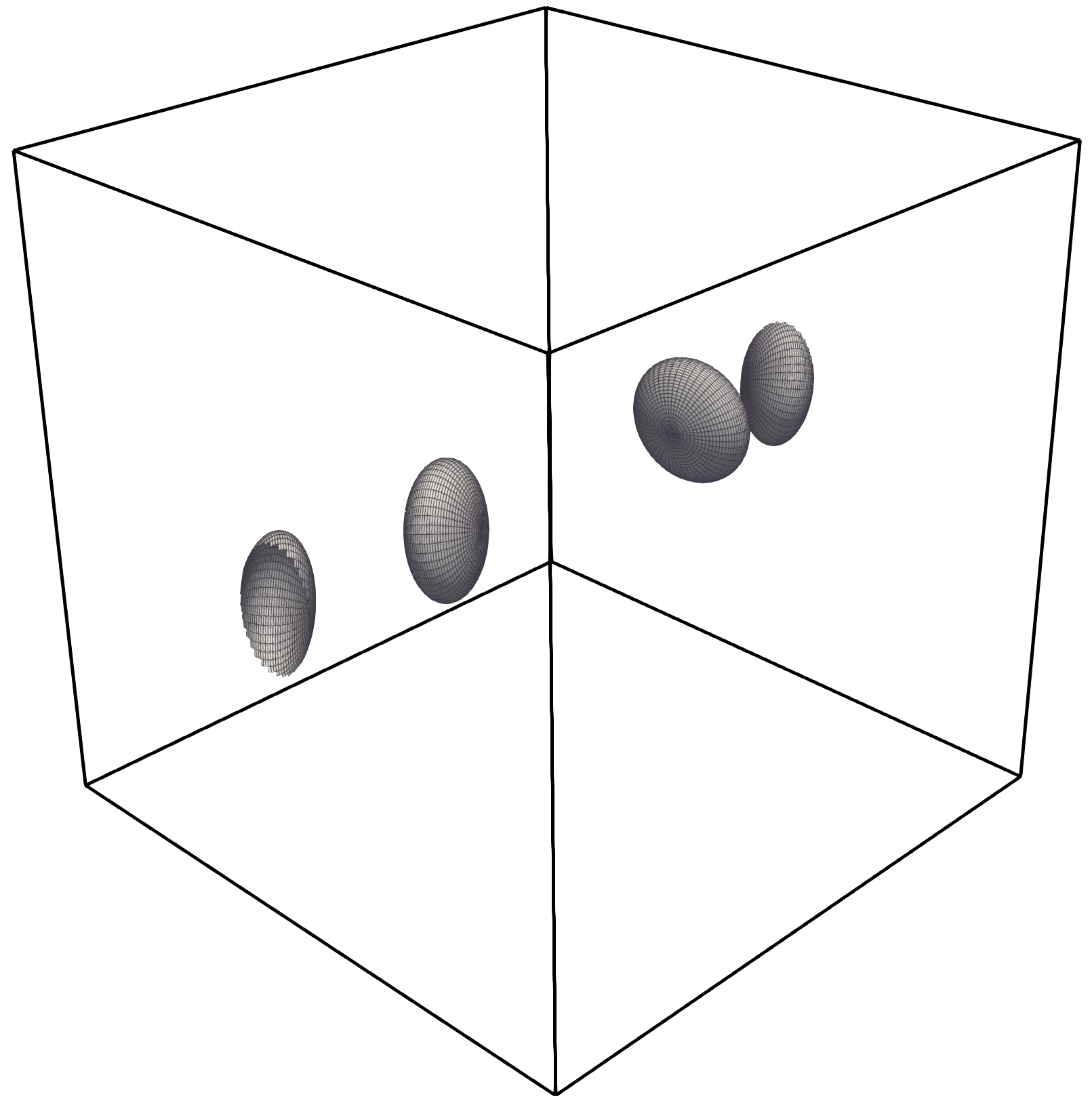}} 
	\hfill
	\subfloat[Reconstructed random structure]{\includegraphics[width=0.32\textwidth]{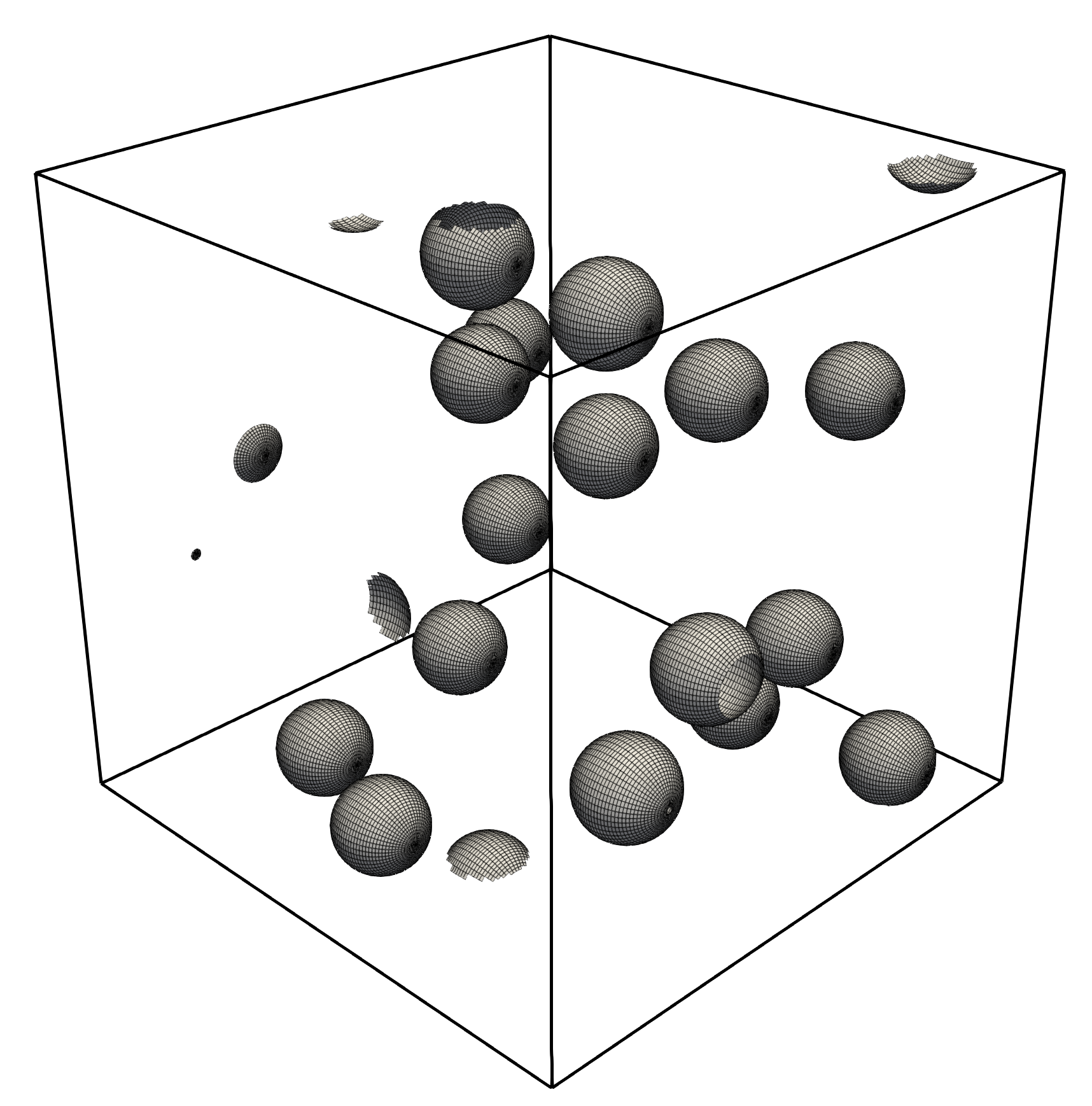}} 
	\caption{Three validation cases for same-size reconstruction. The original (top) and reconstructed (bottom) structures are identical up to point symmetry and translation for volume elements with few inclusions (left and center). This is not the case with many inclusions (right) because of local minima.\label{fig:numexp1}}
\end{figure}

As a first validation, the original and reconstructed microstructure are set to an equal and small size.
As shown in \autoref{fig:numexp1}, the validation structures are composed of three spherical and ellipsoidal inclusions, respectively.

For the first structure shown in \autoref{fig:numexp1} (a), it can be seen that the two right-hand side inclusions in the reconstruction result in \autoref{fig:numexp1} (d) are at the exact same positions as in the original structure.
At a first glance, the position of the left-hand side inclusion seems incorrect, however, it is placed at a symmetrically equivalent position.
More precisely, a point symmetry around the center between the other two ellipsoids moves the inclusion out of the unit cell and under periodic boundary conditions, it re-enters at the observed position.
Similarly, for the second structure in \autoref{fig:numexp1} (b), the reconstruction in \autoref{fig:numexp1} (e) is exact up to a point symmetry around the origin and a displacement.
Because the correlations are intentionally designed to be symmetrical with respect to translation and rotation by $180$°, it can be said that the structures are reconstructed perfectly.
This is confirmed by the corresponding descriptor errors, which are given together with the performance data in \autoref{tab:results}.
However, it should be noted that for more inclusions as in \autoref{fig:numexp1} (c), the result (f) is acceptable but not identical to the reference, as can be seen in terms of the loss in \autoref{tab:results}.
This is further discussed in \autoref{sec:limitationsandscaling}.

\subsection{Application to reduced-size microstructures}
\label{sec:numericalexperiment3}
\begin{figure}[t]
	\centering
	\subfloat[Original random spheres structure]{\includegraphics[width=0.32\textwidth]{figures/wf9_target.png}}
	\hfill
	\subfloat[Original pairwise inclusion structure]{\includegraphics[width=0.32\textwidth]{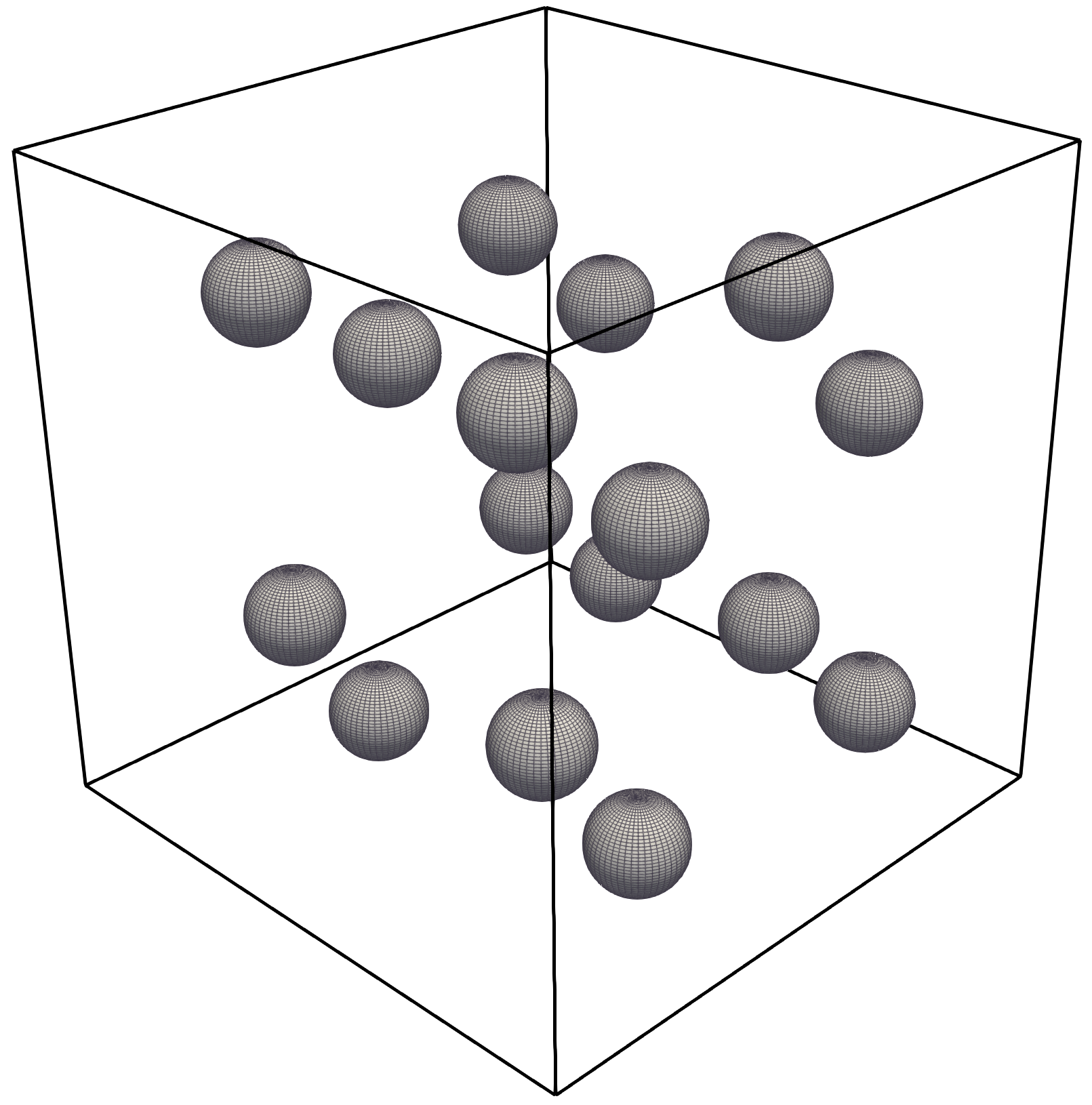}} 
	\hfill
	\subfloat[Original chain structure]{\includegraphics[width=0.32\textwidth]{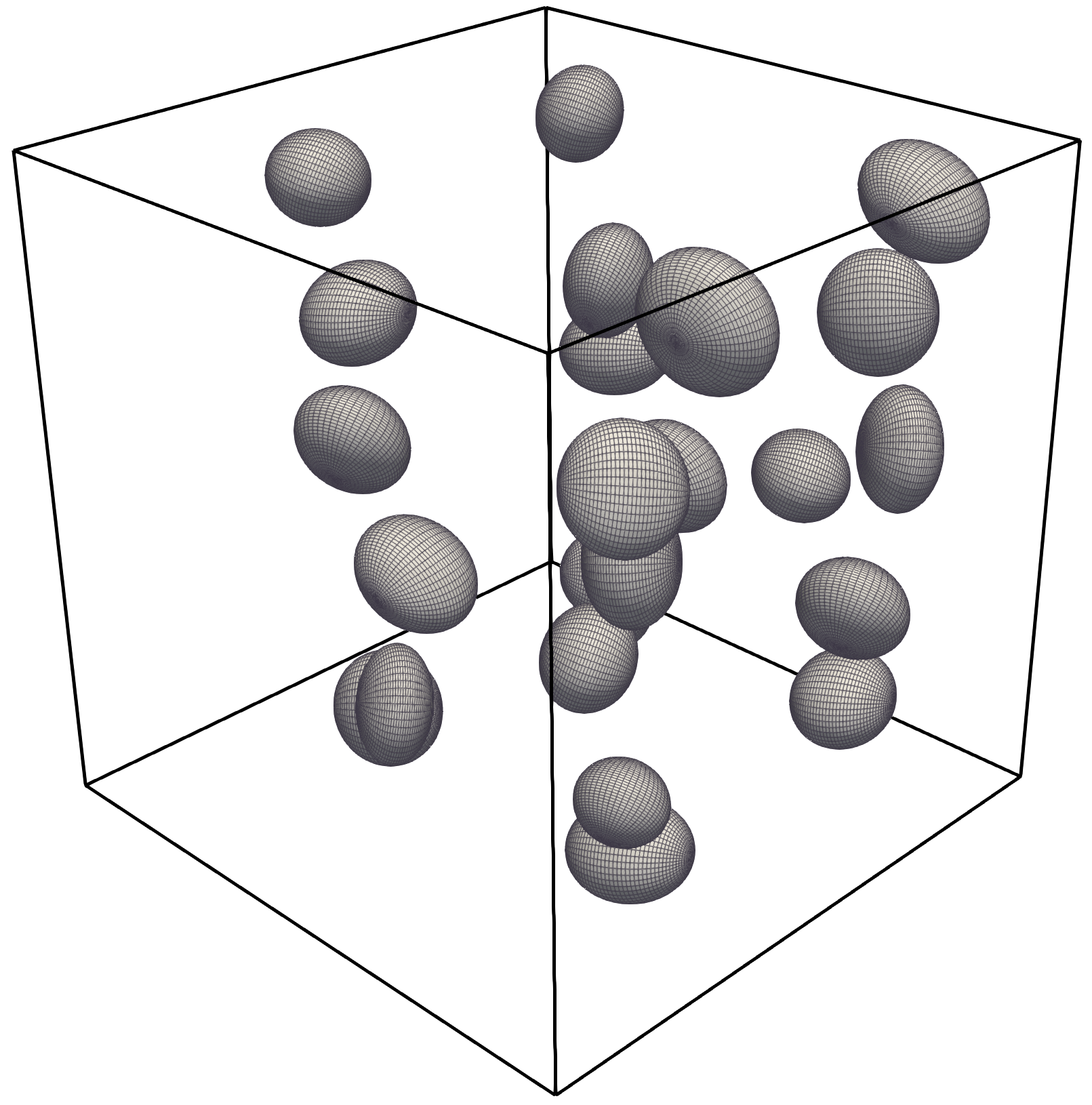}} 
	\vfill
	\subfloat[Reconstructed random structure]{\includegraphics[width=0.32\textwidth]{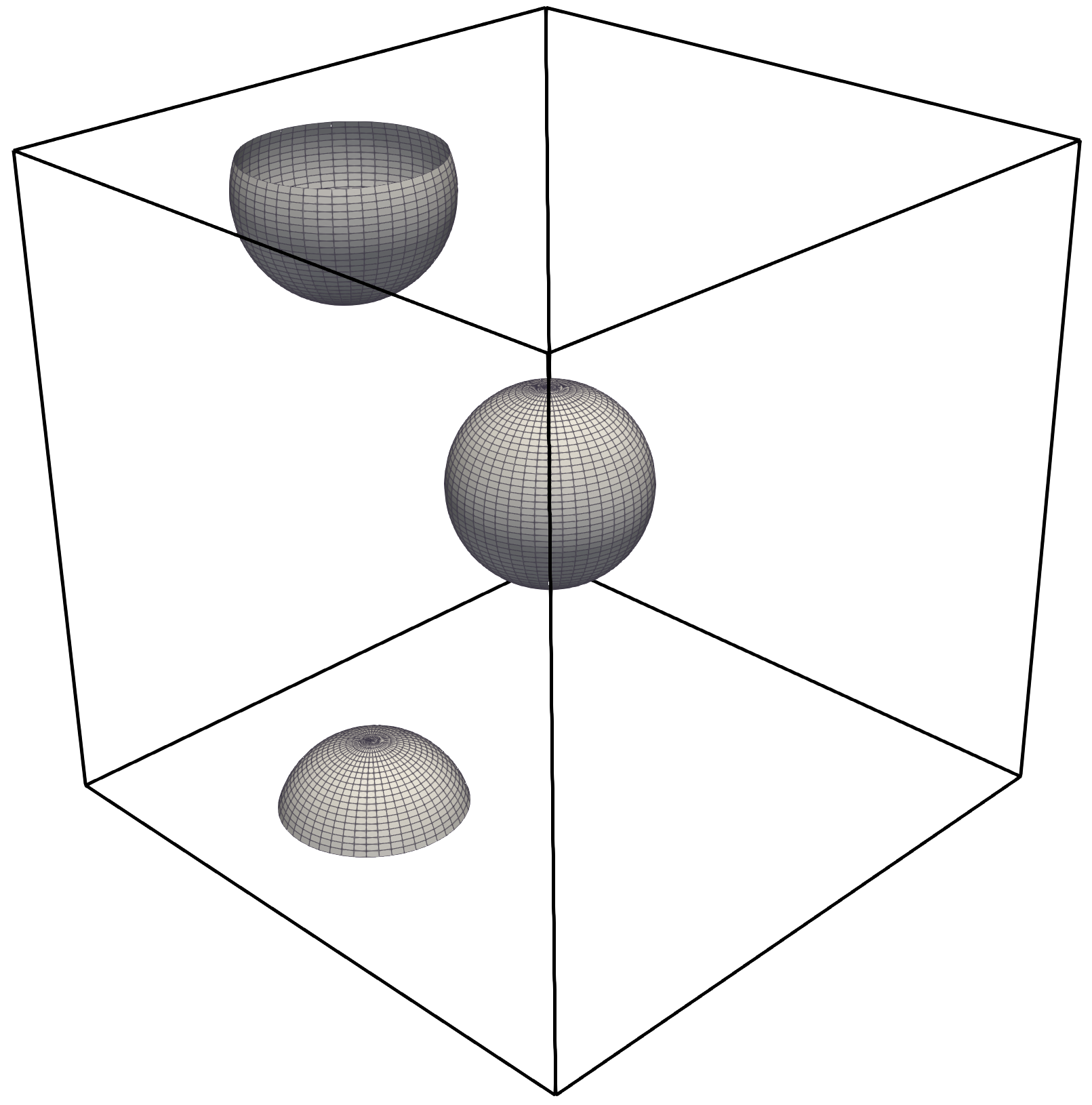}}
	\hfill
	\subfloat[Reconstructed pairwise inclusion structure]{\includegraphics[width=0.32\textwidth]{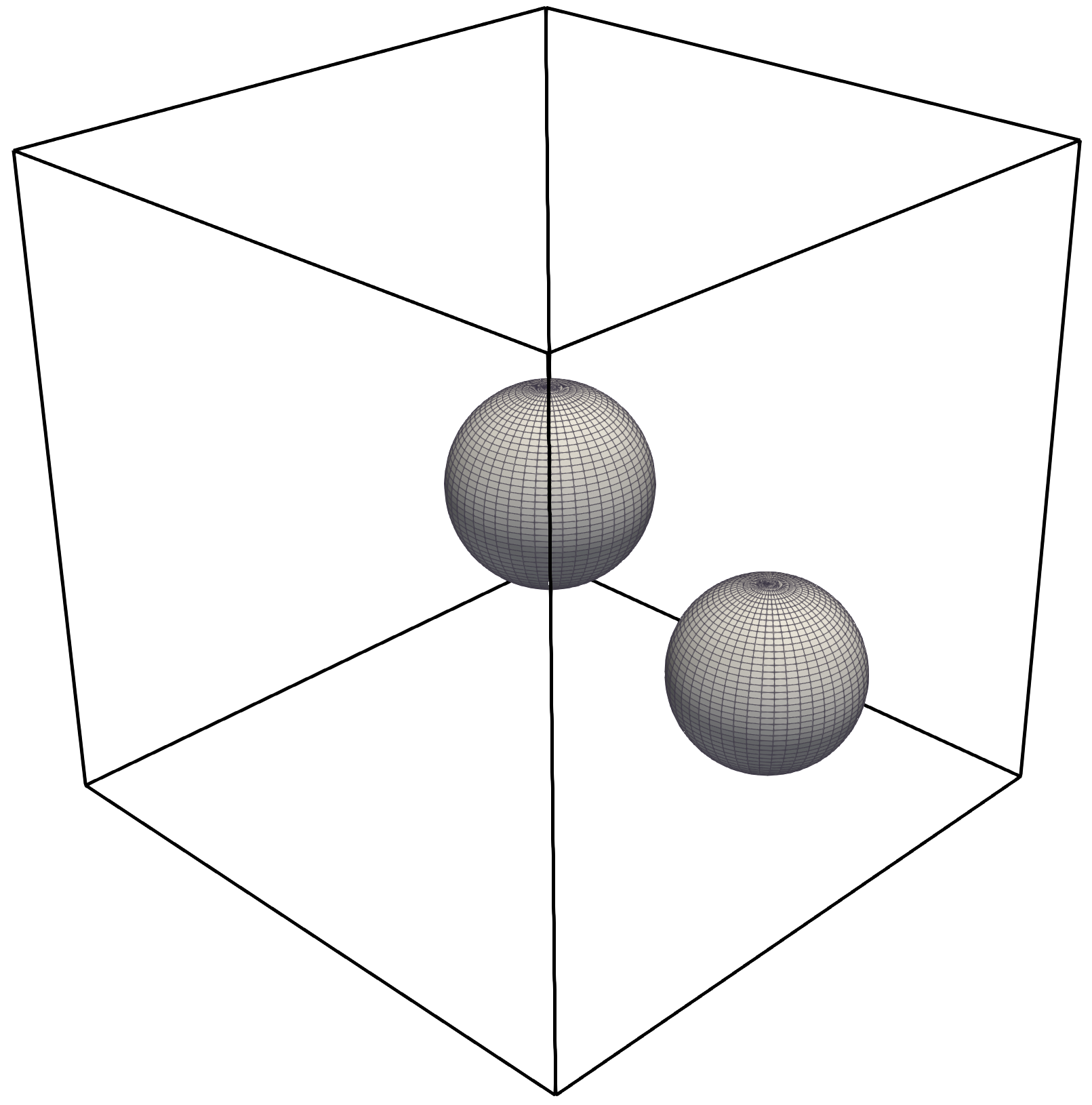}} 
	\hfill
	\subfloat[Reconstructed chain structure]{\includegraphics[width=0.32\textwidth]{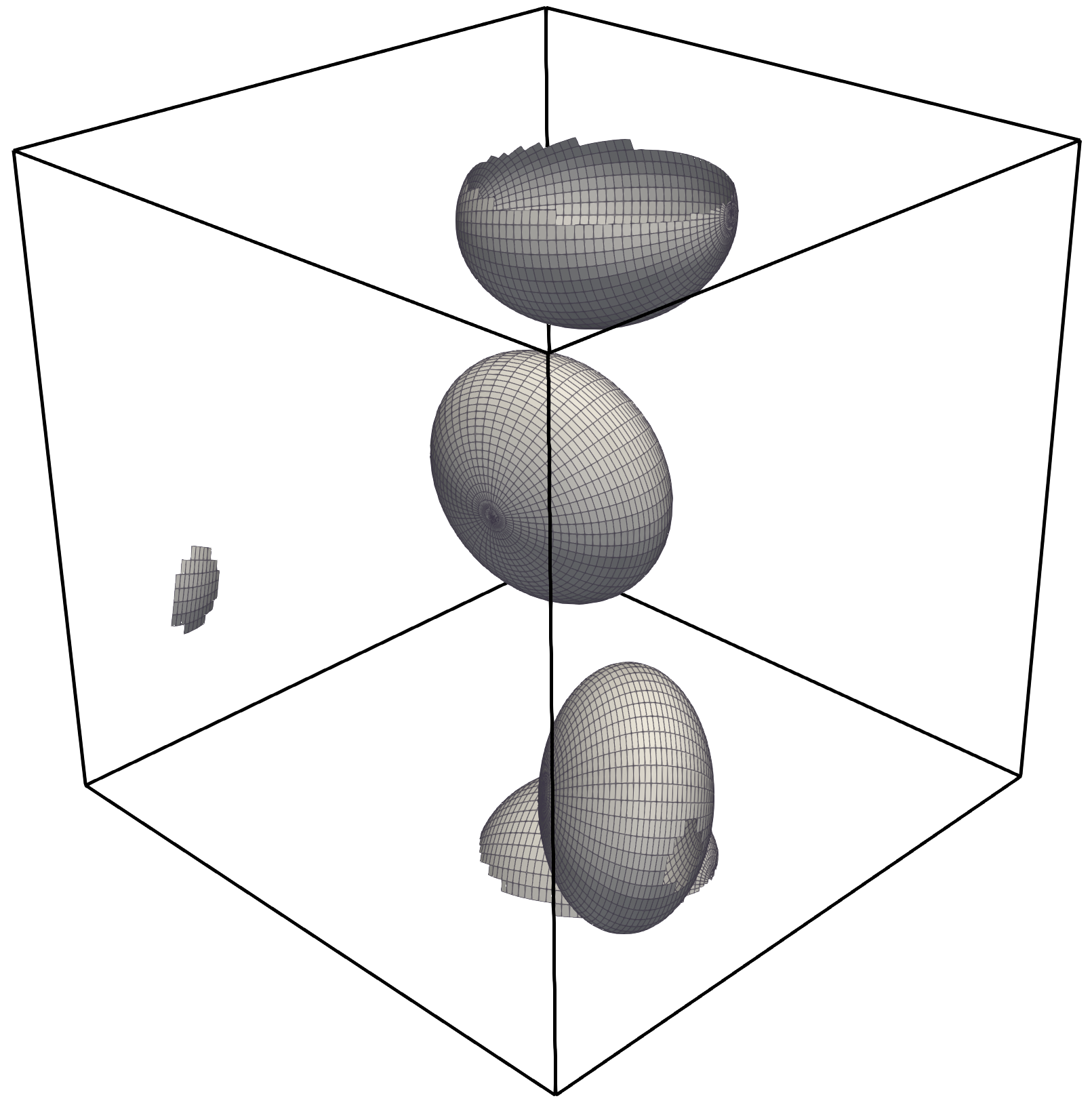}} 
	\caption{Three validation cases for reduced-size reconstruction, i.e. where the reconstruction aims at generating a small, periodic volume element (bottom) from a larger structure (top). The size of the reconstructed volume elements is magnified by a factor of two per direction for visualization purposes. It can be seen that the main characteristics of the original structure are captured well. \label{fig:numexp3}}
\end{figure}

With the method being validated for structures with a small number of inclusions, it is now applied to its intended use-case.
A small and representative periodic structure is reconstructed from a larger domain, which might be a computed tomography scan in real application\footnote{In that case, \emph{Karambola}~\cite{kapfer_karambola_2023} and \emph{MCRpy}~\cite{seibert_microstructure_2022} can be used to compute the desired value of the Minkowski functionals and the two-point correlation, respectively.}.
The reconstructed structure is half as large in each direction, i.e., has an eighth of the volume.
The structures are depicted in \autoref{fig:numexp3} and the corresponding loss values and performance data are given in \autoref{tab:results}.

As a first example, 16 spheres of equal size are placed randomly in a volume as shown in \autoref{fig:numexp3} (a)\footnote{This is identical to \autoref{fig:numexp1} (c).}. 
Because the original volume is eight times larger than the reconstructed structure, observing two inclusions in \autoref{fig:numexp3} (d) meets the expectations.
This is not the case if a domain is randomly sub-sampled from \autoref{fig:numexp3} (a), which would contain between one and three inclusions.

As a second example, in order to verify the plausibility of the placement, \autoref{fig:numexp3} (b) comprises pairs of inclusions where the relative displacement $\ve{r}$ between the centers is drawn at random from uniform distributions~$\mathcal{U}$ with~$r_1 \sim U(0.15, 0.25)$, $r_2 \sim U(-0.25, -0.15)$ and $r_3 \sim U(0.19, 0.39)$.
For the reconstructed structure, $\ve{r} = [0.220, -0.230, 0.311]$ is observed.
Although the elements of $\ve{r}$ lie within the boundaries of the uniform distributions, it is interesting to note that $\ve{r}$ is neither the mean of the distributions, nor of the sampled values\footnote{Not shown here.}.

As a third example, \autoref{fig:numexp3} (c) comprises multiple chains of ellipsoidal inclusions with varying shape and orientation.
Reconstructing inclusions chains is relevant in the context of pre-structured magnetorheological elastomers, where they are created by applying an external magnetic field during solidification.
The metallic inclusions allow for switching their macroscopic behavior by applying an external magnetic field.
This property enables applications to various engineering problems such as actuators and sensors \cite{tian_sensing_2011,volkova_motion_2016} or dampers \cite{carlson_mr_2000,li_highly_2013}.
As expected, a chain is reconstructed that may be used for efficient simulations with a small domain and periodic boundary conditions.
Thus, while extensive research has been conducted on the modeling and simulation \cite{bustamante_numerical_2011,kalina_multiscale_2020,metsch_field-induced_2020,mukherjee_microstructurally-guided_2019,rambausek_computational_2022} as well as experimental and design aspects \cite{bastola_review_2020,danas_experiments_2012,lokander_improving_2003} of MREs, a tailored reconstruction algorithm may enable a stronger interaction between these fields to bring MREs closer to practical application. 
\begin{table*}[b]
	\centering
	\caption{Loss function and wallclock time for all reconstructions carried out in the present work. The wallclock time is almost identical to the duration of the inclusion placement, since the computational cost of the other steps in \autoref{alg:alg} can be neglected in practice.}
	\label{tab:results}
	\begin{tabular}{c  c  c  c  c  c  }
\toprule
ID & Reconstructed from & Size & Visualized in & $\lossfun^\mathrm{\elepos}$ & Time in s \\
\midrule
1 & three spheres & same & \autoref{fig:numexp1} (d) & $7.3 \cdot 10^{-31}$ & $19.5$ \\
2 & three ellipsoids & same & \autoref{fig:numexp1} (e) & $6.5 \cdot 10^{-31}$ & $99.3$ \\
3 & random & same & \autoref{fig:numexp1} (f) & $1.7 \cdot 10^{-1}$ & $195.4$ \\
4 & random & reduced & \autoref{fig:numexp3} (d) & $2.5 \cdot 10^{1}$ & $35.1$ \\
5 & pairwise inclusion & reduced & \autoref{fig:numexp3} (e) & $8.5 \cdot 10^{0}$ & $18.7$ \\
6 & chain & reduced & \autoref{fig:numexp3} (f) & $1.1 \cdot 10^{1}$ & $196.9$ \\
\bottomrule
	\end{tabular}
\end{table*}

\subsection{Scalability and limitations}
\label{sec:limitationsandscaling}
Because the degrees of freedom of the inclusions are considered directly instead of discretizing them to a voxel grid or tesselating the surface, the dimensionality of the optimization problem is independent of the volume size or resolution, but rather grows linearly with the number of inclusions.
The cost of reconstructing ellipsoid shapes and orientations from Minkowski tensor distributions scales linearly in the number of inclusions and can be neglected when compared to the inclusion placement.
In this context, the computational cost of evaluating the loss function scales as the spatial correlations and the surface-to-surface distance computation.
In the current implementation, distances are computed between all ellipsoids, making this term scale as~$\mathcal{O}((N^\mathrm{incl})^2)$, but a verlet list or a similar data structure could be used to reduce this number.
The distance computation cannot be discarded to save computational effort, because avoiding collisions is not only a physical requirement in the materials of interest, but also a mathematical requirement for the validity of the presented analytical solutions.
In order to compute the Fourier coefficients of the indicator function,~$\mathcal{O}(\nfourier{1}\nfourier{2}\nfourier{3})$ coefficients need to be computed for each of the~$N^\mathrm{incl}$ ellipsoids.
The Fourier transform in \autoref{eq:fouriertransforms2} to obtain $\approximate{\tpc}^{\rveellipsuperscript}$ from $|\fourierindicator{jkl}^{\rveellipsuperscript} |^2$ scales as~$\mathcal{O}(\nfourier{1} \nfourier{2} \nfourier{3} \log \nfourier{1} \log \nfourier{2} \log \nfourier{3})$.
In principle, one might equally well use $|\fourierindicator{jkl}^{\rveellipsuperscript} |^2$ as a descriptor in the placement loss function \autoref{eq:lossfuncp} to save computational effort in future works.
However, the computational cost of evaluating the loss function does not provide a good estimate for actually solving the optimization problem in practice because local minima become more pronounced as~$N^\mathrm{incl}$ increases and significantly more iterations are needed.
In fact, while efficiently reconstructing a small volume element from a larger reference with inclusion chains as in \autoref{fig:numexp3} (c) is a strength of the method, an equally large volume element cannot be reconstructed at all in a feasible time because of local minima.
Hence, if this is of interest, a suitable or even tailored optimizer is required or the formulation of the optimization problem needs to be adapted.

\begin{figure}[t]
	\centering
	\subfloat[Strongly non-ellipsoidal inclusion]{\includegraphics[width=0.32\textwidth]{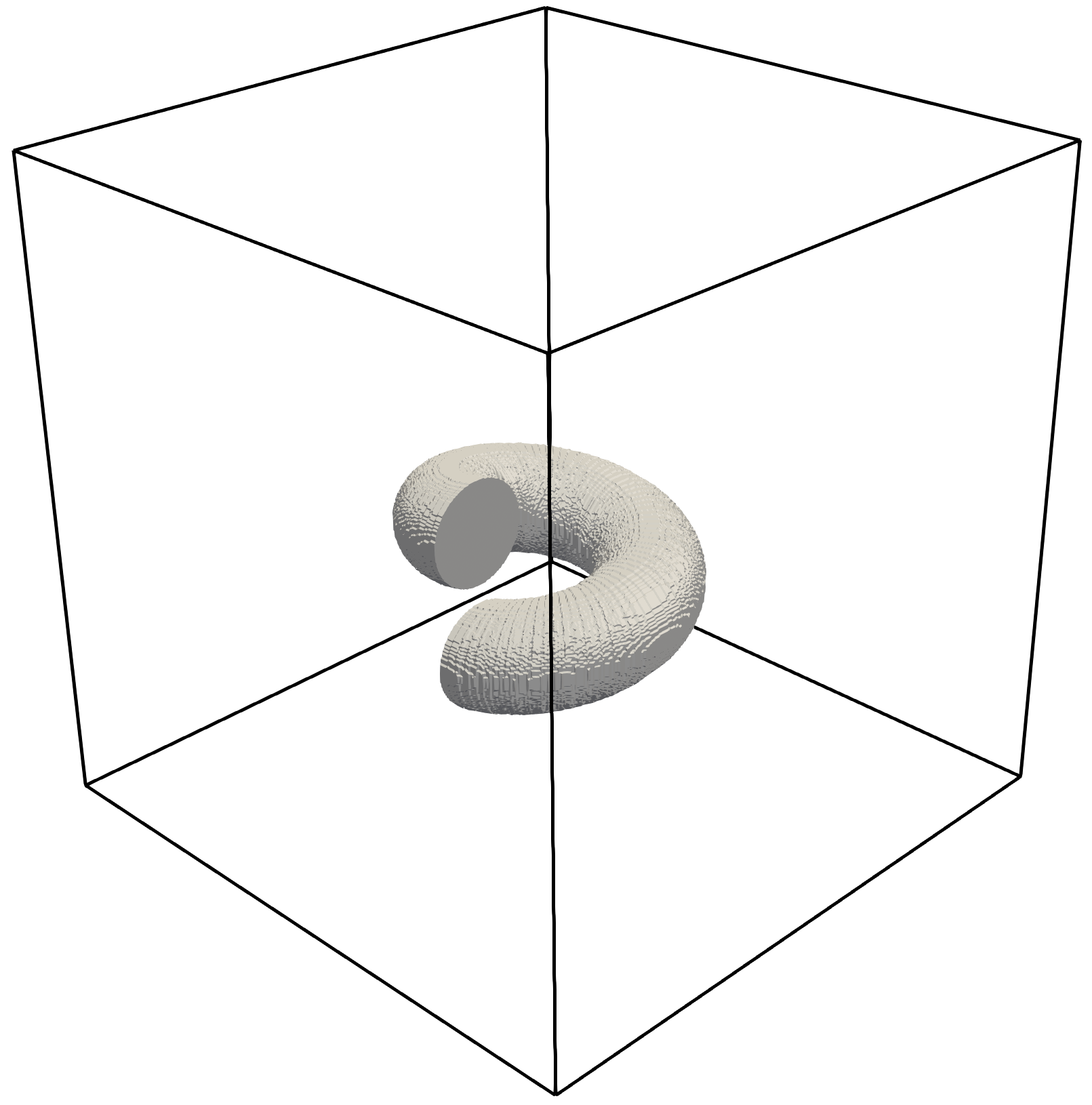}}
	\hfill
	\subfloat[Reconstruction from (a)]{\includegraphics[width=0.32\textwidth]{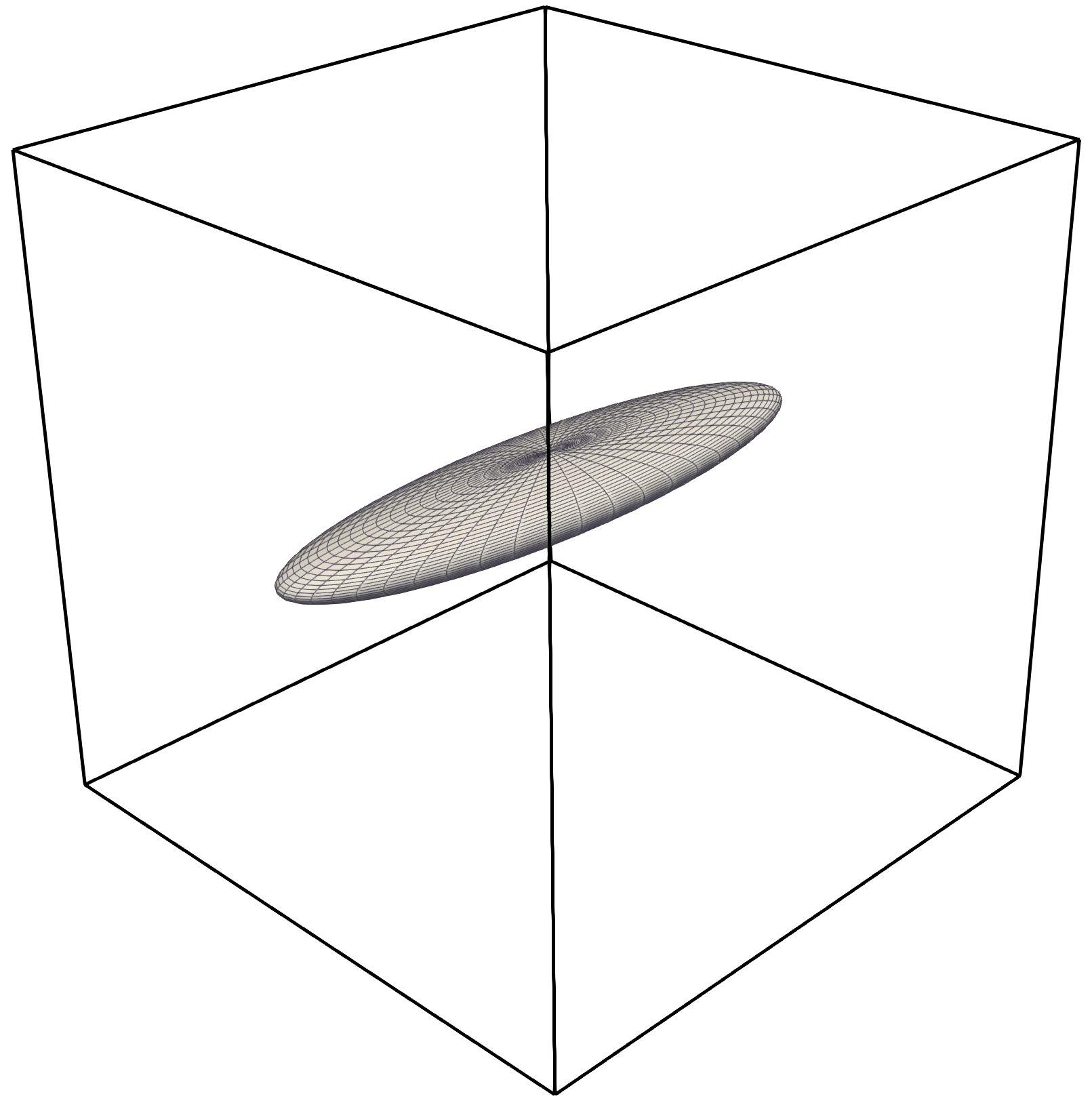}} 
	\caption{Investigation of the reconstruction of strongly non-ellipsoidal inclusions. Because Minkowski functionals are used for the reconstruction, the approximation of the shape in (a) is not optimized in terms of the spatial extent, but rather in terms of volume, surface area and other captures characteristics.\label{fig:torus}}
\end{figure}

A second limitation naturally lies in the shape of the inclusions, which is approximated as ellipsoidal in this work.
If the considered inclusions significantly differ, such as the spiral in \autoref{fig:torus}~(a), it should be noted that they are not approximated in terms of spatial extent, but in terms of Minkowski functionals. 
Since the first two scalar Minkowski functionals coincide with the volume fraction and the surface area of the inclusion, non-convex inclusions can only be approximated by very flat ellipsoids.
Based on this observation, the orientation and the exact aspect ratios are determined by the remaining functionals and the reader is referred to~\cite{schroder-turk_tensorial_2010} for an interpretation of the same.
The closest possible ellipsoid to \autoref{fig:torus}~(a) in terms of Minkowski funcionals is shown in \autoref{fig:torus}~(b) and demonstrates that \autoref{fig:torus}~(a) clearly exceeds the range of applicability of the method.
Although complex inclusions can be approximated very well by overlapping simple geometric shapes such as spheres in~\cite{nakka_computationally_2022}, this would later need to be considered in deriving global descriptors such as the two-point correlation. 
As an alternative solution, other inclusion shape descriptors might be considered, such as the cluster correlation function~\cite{jiao_superior_2009}.
It can be derived analytically based on the presented results by replacing $\fourierindicator{jkl}^{\rveellipsuperscript}$ in \autoref{eq:fouriertransforms2} by $\fourierindicator{jkl}^{\ellipsoidsuperscript}$.
This is left for future investigations since strongly non-ellipsoidal inclusions lie beyond the scope of this work.

In summary, while the method is highly effective and accurate for reconstructing a small number of ellipsoidal inclusions, further developments are needed for cases where these inclusions are not actually ellipsoidal or are very numerous.

\section{Conclusions}
\label{sec:summary}
In this work, a tailored reconstruction algorithm is presented for microstructures with ellipsoidal inclusions like magnetorheological elastomers.
Unlike in previous work, these inclusions are non-overlapping. 
Building upon gradient-based microstructure reconstruction as in \emph{MCRpy}~\cite{seibert_microstructure_2022} and inspired by the low-dimensional microstructure approximation in the work of Scheunemann et al.~\cite{scheunemann_design_2015,scheunemann_scale-bridging_2017}, a high efficiency is reached.
This is rooted in four main features of the method, namely (i) the low dimensionality of the minimization problem, (ii) the subdivision of the optimization into consecutive stages, (iii) the computational efficiency of the derived analytical solutions for the descriptors and (iv) the differentiability of the same, allowing for gradient-based optimization.

For this, analytical expressions are derived to compute the Minkowski functionals and the spatial two-point correlation\footnote{Although not used, the cluster correlation function can be obtained trivially from this derivation.} as a function of the microstructure parametrization, i.e. the ellipsoid semiaxes, orientations and positions.
Based on these descriptors, an algorithm is presented that sequentially samples individual tuples of Minkowski functionals, recovers the corresponding inclusion shapes and orientations and finally places all ellipsoids in the volume element.
This method is validated by means of various numerical experiments.
It is concluded that the method is very well suited for efficiently generating small statistical volume elements from larger structures, however, a transparent summary of the method's limitations is also given in the following.

By directly representing inclusions analytically, the computational cost of the method is independent of the structure resolution, however, it does scale with the number of inclusions.
Although the cost function for determining the ellipsoid location can still be evaluated efficiently for ten or more inclusions, the optimization problem itself scales very unfavorably, as the increasing number of local minima overstrains the utilized global optimizer.
Besides that, although analytically deriving descriptors as functions of the microstructure parametrization yields remarkable efficiency, this derivation can also represent a hurdle when new descriptors are to be used\footnote{Note in this context the ambiguity of the two-point correlation function~\cite{fernandez_generation_2020} and the comparatively low effort of adding descriptors to \emph{MCRpy}~\cite{seibert_microstructure_2022}.}.
Finally, it is observed that the nature of the Minkowski functionals leads to unexpected behaviour if the inclusions to reconstruct are strongly non-ellipsoidal.

For future work, besides the derivation of further descriptors, it should also be investigated how well the limit case of elongated ellipsoids can be used to approximate short fibers or if the descriptor derivations would need to be repeated for a different geometry.
Furthermore, it would be promising to develop a microstructure descriptor to quantify the distribution of Minkowski functionals instead of individual inclusions.
As discussed previously, this could be achieved by a kernel density estimate.
Finally, reconstructing 3D structures from 2D slices would enable a plethora of applications.
However, such an extension cannot be performed as easily as in voxel- and slice-based methods~\cite{zhang_slice--voxel_2021,kench_generating_2021,seibert_descriptor-based_2022} because 3D descriptors need to be corrected for the fact that inclusions are cut at random~\cite{anderson_automated_2023}.

\section*{Acknowledgements}
The group of M. Kästner thanks the German Research Foundation DFG which supported this work under Grant number KA 3309/18-1.
The authors are grateful to the Centre for Information Services and High Performance Computing [Zentrum für Informationsdienste und Hochleistungsrechnen (ZIH)] TU Dresden for providing its facilities for high throughput calculations.

\section*{Competing interest statement}
The authors declare no competing interest.

\section*{Data availability statement}
The code and data is made available by the authors upon reasonable request.

\printcredits

\appendix

\section{Derivation of analytical expressions for Minkowski functionals}
\label{sec:appendixderivations}
In order to calculate analytical expressions for Minkowski functionals of ellipsoids, the expressions of the surface element~$\surfele{3}$, the unit normal vector~$\ve{\normal}$ and the mean curvature~$\meancurve$ are derived in the following.
The covariant basis vectors are defined as
\begin{align}
\covbasisvec{\polarangle}\surfacepara&=\frac{\partial \ve{\position}}{\partial \polarangle}=-\semiax_1 \sin(\azimutangle) \sin(\polarangle) \unitvec{1} +\semiax_2 \sin(\azimutangle) \cos(\polarangle)\unitvec{2} \quad \text{and}\\
\covbasisvec{\azimutangle}\surfacepara&=\frac{\partial \ve{\position}}{\partial \azimutangle}=\semiax_1 \cos(\azimutangle) \cos(\polarangle) \unitvec{1}+\semiax_2 \cos(\azimutangle) \sin(\polarangle)\unitvec{2}-\semiax_3 \sin(\azimutangle)\unitvec{3}.
\end{align}
and allow computing a surface element $\surfele{3}$ of the ellipsoid surface ~\cite{carmo_differential_2016} as
\begin{equation}
\label{eq:surface-element}
\surfele{3}\surfacepara=\sqrt{\fffE\surfacepara \fffG\surfacepara-\fffF^2\surfacepara} \diff \azimutangle \diff \polarangle.
\end{equation}
using the coefficients
\begin{align}
\fffE\surfacepara&=\covbasisvec{\polarangle}\surfacepara\cdot \covbasisvec{\polarangle}\surfacepara=\sin^2(\azimutangle) \left[(\semiax_1)^2 \sin^2(\polarangle)+(\semiax_2)^2 \cos^2(\polarangle)\right],\\
\fffF\surfacepara&=\covbasisvec{\azimutangle}\surfacepara\cdot \covbasisvec{\polarangle}\surfacepara=\left[(\semiax_2)^2-(\semiax_1)^2\right] \cos(\azimutangle)\sin(\azimutangle)\cos(\polarangle)\sin(\polarangle) \quad \text{and}\\
\fffG\surfacepara&=\covbasisvec{\azimutangle}\surfacepara\cdot \covbasisvec{\azimutangle}\surfacepara=\cos^2(\azimutangle) \left[(\semiax_1)^2 \cos^2(\polarangle)+(\semiax_2)^2 \sin^2(\polarangle)\right]+(\semiax_3)^2 \sin^2(\azimutangle) \; .
\end{align}
The unit normal vector $\ve{\normal}$ on the ellipsoid surface is defined as 
\begin{equation}
\label{eq:normal-ellip-def}
\ve{\normal}\surfacepara=
\frac{
\covbasisvec{\polarangle}\surfacepara \times \covbasisvec{\azimutangle}\surfacepara
}
{
\norm{
\covbasisvec{\polarangle}\surfacepara \times \covbasisvec{\azimutangle}\surfacepara
}
} \; ,
\end{equation}
where the cross product of the covariant basis vectors is
\begin{equation}
\label{eq:cross-prod}
\covbasisvec{\polarangle}\surfacepara \times \covbasisvec{\azimutangle}\surfacepara=
-\semiax_2\semiax_3 \sin^2(\azimutangle) \cos(\polarangle) \unitvec{1}
-\semiax_1\semiax_3 \sin^2(\azimutangle) \sin(\polarangle) \unitvec{2}
-\semiax_1 \semiax_2 \sin(\azimutangle) \cos(\azimutangle) \unitvec{3}
\end{equation}
with the Euclidean norm
\begin{align}
\label{eq:norm-cross-prod}
\norm{
\covbasisvec{\polarangle}\surfacepara \times \covbasisvec{\azimutangle}\surfacepara
}&=
\sin(\azimutangle)
\sqrt{
(\semiax_2)^2(\semiax_3)^2 \sin^2(\azimutangle) \cos^2(\polarangle)
+(\semiax_1)^2(\semiax_3)^2 \sin^2(\azimutangle) \sin^2(\polarangle)
+(\semiax_1)^2(\semiax_2)^2 \cos^2(\azimutangle)
}
\\
&=\sin(\azimutangle)
\sqrt{
\sin^2(\polarangle) \auxmeancurve{1}(\azimutangle)
+\cos^2(\polarangle)\auxmeancurve{2}(\azimutangle)
}
=\sqrt{\fffE\surfacepara\fffG\surfacepara-\fffF^2\surfacepara} \; ,
\end{align}
and the auxiliary functions
\begin{align}
\auxmeancurve{1}(\azimutangle)&=(\semiax_1)^2 \left[(\semiax_3)^2\sin^2(\azimutangle)+(\semiax_2)^2\cos^2(\azimutangle)\right] \quad \text{and}\\
\auxmeancurve{2}(\azimutangle)&=(\semiax_2)^2 \left[(\semiax_3)^2\sin^2(\azimutangle)+(\semiax_1)^2\cos^2(\azimutangle)\right] \; .
\end{align}
Substituting \autoref{eq:cross-prod} and \autoref{eq:norm-cross-prod} into \autoref{eq:normal-ellip-def} yields the required expression for the unit normal vector
\begin{align}
\ve{\normal}\surfacepara &=
\normal_1 \unitvec{1}+\normal_2 \unitvec{2}+\normal_3 \unitvec{3}
\\
&= \frac{1}
{\sqrt{\fffE\surfacepara\fffG\surfacepara-\fffF^2\surfacepara}}
\left(-\semiax_2\semiax_3 \sin^2(\azimutangle) \cos(\polarangle) \unitvec{1}
-\semiax_1\semiax_3 \sin^2(\azimutangle) \sin(\polarangle) \unitvec{2}
-\semiax_1 \semiax_2 \sin(\azimutangle) \cos(\azimutangle) \unitvec{3}\right).
\label{eq:ellip-normal-final-def}
\end{align}

The derivatives of the covariant basis vectors with regards to the coordinates $\azimutangle$ and $\polarangle$
\begin{align}
\covbasisvec{\polarangle,\polarangle}\surfacepara
&=\frac{\partial \covbasisvec{\polarangle}\surfacepara}{\partial \polarangle}=
-\semiax_1 \sin(\azimutangle) \cos(\polarangle) \unitvec{1} -\semiax_2 \sin(\azimutangle) \sin(\polarangle)\unitvec{2},\\
\covbasisvec{\azimutangle,\polarangle}\surfacepara&=\covbasisvec{\polarangle,\azimutangle}\surfacepara
=\frac{\partial \covbasisvec{\azimutangle}\surfacepara}{\partial \polarangle}
=\frac{\partial \covbasisvec{\polarangle}\surfacepara}{\partial \azimutangle}
=
-\semiax_1 \cos(\azimutangle) \sin(\polarangle) \unitvec{1} +\semiax_2 \cos(\azimutangle) \cos(\polarangle)\unitvec{2} \quad \text{and}\\
\covbasisvec{\azimutangle,\azimutangle}\surfacepara
&=\frac{\partial \covbasisvec{\azimutangle}\surfacepara}{\partial \azimutangle}=
-\semiax_1 \sin(\azimutangle) \cos(\polarangle) \unitvec{1} -\semiax_2 \sin(\azimutangle) \sin(\polarangle)\unitvec{2}-\semiax_3 \cos(\azimutangle)\unitvec{3} 
\end{align}
define the coefficients of the second fundamental form of differential geometry
\begin{align}
\sffe\surfacepara &=\ve{\normal}\surfacepara \cdot \covbasisvec{\polarangle,\polarangle}\surfacepara=
\frac{
\semiax_1 \semiax_2 \semiax_3 \sin^3(\azimutangle)
}
{
\sqrt{\fffE\surfacepara\fffG\surfacepara-\fffF^2\surfacepara}
},\\
\sfff\surfacepara &=\ve{\normal}\surfacepara \cdot \covbasisvec{\azimutangle,\polarangle}\surfacepara=
0\quad \text{and}\\
\sffg\surfacepara &=\ve{\normal}\surfacepara \cdot \covbasisvec{\azimutangle,\azimutangle}\surfacepara=
\frac{
\semiax_1 \semiax_2 \semiax_3 \sin(\azimutangle)
}
{
\sqrt{\fffE\surfacepara\fffG\surfacepara-\fffF^2\surfacepara}
},
\end{align}
which can be used to compute the mean curvature $\meancurve$ as
\newcommand{\meancurveeq}{\frac{
\auxmeancurve{3}(\azimutangle)\sin^2(\polarangle)+\auxmeancurve{4}(\azimutangle)\cos^2(\polarangle)}
{\left[\auxmeancurve{1}(\azimutangle)\sin^2(\polarangle)+\auxmeancurve{2}(\azimutangle)\cos^2(\polarangle)\right]^{\frac{3}{2}}}}
\begin{equation}
\label{eq:ellip-meancurve}
\meancurve\surfacepara=\frac{1}{2}(\princurve{1}\surfacepara+\princurve{2}\surfacepara)=
\frac{1}{2}
\frac{\sffe\surfacepara\fffG\surfacepara-2\sfff\surfacepara\fffF\surfacepara+\sffg\surfacepara\fffE\surfacepara}{\fffE\surfacepara\fffG\surfacepara-\fffF^2\surfacepara}
=\frac{\semiax_1 \semiax_2 \semiax_3}{2}
\meancurveeq,
\end{equation}
with the auxiliary functions
\begin{align}
\auxmeancurve{3}(\azimutangle)&=\left[(\semiax_1)^2+(\semiax_3)^2\right]\sin^2(\azimutangle)+\left[(\semiax_1)^2+(\semiax_2)^2\right]\cos^2(\azimutangle) \quad \text{and}\\
\auxmeancurve{4}(\azimutangle)&=\left[(\semiax_2)^2+(\semiax_3)^2\right]\sin^2(\azimutangle)+\left[(\semiax_1)^2+(\semiax_2)^2\right]\cos^2(\azimutangle).
\end{align}

With the definition of the ellipsoid surface in \autoref{eq:ellip-surface}, the surface element in \autoref{eq:surface-element}, the normal vector in \autoref{eq:ellip-normal-final-def} and the mean curvature in \autoref{eq:ellip-meancurve}, the tensorial Minkowski functionals of an ellipsoid can be calculated.

\bibliographystyle{cas-model2-names}

\bibliography{main}

\end{document}